\documentclass[reprint,amsfonts, amssymb, amsmath,  showkeys,pra, superscriptaddress, nofootinbib, twocolumn,longbibliography]{revtex4-2}

\usepackage{xcolor}

\usepackage{float}
\makeatletter
\let\newfloat\newfloat@ltx
\makeatother
\usepackage[english]{babel}
\usepackage[utf8]{inputenc}
\usepackage{graphics}
\usepackage{selinput}
\usepackage[normalem]{ulem}
\usepackage[shortlabels]{enumitem}

\usepackage{braket}
\usepackage{amsthm}
\usepackage{mathtools}
\usepackage{physics}
\usepackage{graphicx}
\usepackage[left=16mm,right=16mm,top=35mm,columnsep=15pt]{geometry} 
\usepackage{adjustbox}
\usepackage{placeins}
\usepackage[T1]{fontenc}
\usepackage{lipsum}
\usepackage{csquotes}
\usepackage{bm}
\usepackage{mathrsfs}

\usepackage[linesnumbered,ruled,vlined]{algorithm2e}
\SetKwInput{kwInit}{Init}

\def\HC{\mathcal{H}}

\newcommand{\fsnull}[1]{}
\newcommand{\old}[1]{}

\usepackage[makeroom]{cancel}
\usepackage[toc,page]{appendix}
\usepackage[colorlinks=true,citecolor=blue,linkcolor=magenta]{hyperref}

\usepackage{tikz}
\tikzset{every picture/.style=remember picture}

\usepackage[utf8]{inputenc}
\usepackage{graphicx}
\usepackage{xcolor}
\usepackage{amsmath}
\usepackage{amsthm}
\usepackage{bm}
\usepackage{bbm}
\usepackage{comment}
\usepackage{appendix}
\usepackage{mathdots}
\usepackage{lipsum}
\usepackage{verbatim}
\usepackage{natbib}
\usepackage{nccmath}
\usepackage{amsfonts}

\newcommand{\poly}{\operatorname{poly}}

\newcommand{\BC}{\mathcal{B}}
\newcommand{\CC}{\mathcal{C}}
\newcommand{\CS}{\mathscr{C}}
\newcommand{\DC}{\mathcal{D}}

\newcommand{\NC}{\mathcal{N}}
\newcommand{\OC}{\mathcal{O}}

\newcommand{\SC}{\mathcal{S}}

\renewcommand{\geq}{\geqslant}
\renewcommand{\leq}{\leqslant}

\renewcommand{\vec}[1]{\boldsymbol{#1}}  %

\newcommand*{\id}{\openone}

\newcommand{\bs}{\textsf{BS}}

\newcommand{\io}{\iota }

\def\C{\mathbb{C}}

\newcommand{\losalamos}{Theoretical Division, Los Alamos National Laboratory, Los Alamos, New Mexico 87545, USA}
\newcommand{\ibm}{IBM Research, Chicago, IL 60606, USA}

\def\be{\begin{equation}}
\def\ee{\end{equation}}
\def\bs{\begin{split}}
\def\e{\end{split}}
\def\ba{\begin{eqnarray}}
\def\bea{\begin{eqnarray}}

\def\tea{\end{eqnarray}}
\def\ea{\end{eqnarray}}
\def\eea{\end{eqnarray}}

\usepackage{amssymb}
\usepackage{dsfont}

\def\be{\begin{equation}}
\def\te{\end{equation}}
\def\ee{\end{equation}}
\def\ba{\begin{eqnarray}}
\def\bea{\begin{eqnarray}}

\def\tea{\end{eqnarray}}
\def\ea{\end{eqnarray}}
\def\eea{\end{eqnarray}}

\begin{document}

\title{Quantum Gaussian processes for prediction of channel observations}

\author{Jonas Jäger}
\affiliation{Department of Computer Science and Institute of Applied Mathematics, University of British Columbia, Vancouver, V6T 1Z4 B.C., Canada}
\affiliation{Stewart Blusson Quantum Matter Institute, Vancouver, V6T 1Z4 B.C., Canada}

\author{Yaroslav Khmelnitskiy}
\affiliation{Institute of Theoretical Physics, Jagiellonian University, Krak\'ow, Poland.}

\author{Paolo Braccia}
\affiliation{\losalamos}
\affiliation{\ibm}

\author{Artur Miroszewski}
\affiliation{$\Phi$-lab, European Space Agency (ESA/ESRIN), Frascati, Italy}
\affiliation{Institute of Theoretical Physics, Jagiellonian University, Krak\'ow, Poland.}
\affiliation{Mark Kac Center for Complex Systems Research, Jagiellonian University, Krak\'ow, Poland}

\author{Diego Garc\'ia-Mart\'in }
\affiliation{Department for Quantum Information and Computation at Kepler (QUICK),\\ Johannes Kepler University, Linz, Austria }

\author{M. Cerezo}
\thanks{cerezo@lanl.gov}
\affiliation{Information Sciences, Los Alamos National Laboratory, Los Alamos, NM 87545, USA}

\author{Piotr Czarnik}
\affiliation{Institute of Theoretical Physics, Jagiellonian University, Krak\'ow, Poland.}
\affiliation{Mark Kac Center for Complex Systems Research, Jagiellonian University, Krak\'ow, Poland}

\begin{abstract}
Given a set of input states, we consider the task of predicting the expectation value of a Pauli observable at the output of an unknown quantum evolution, using only a limited number of measurements. Recently, quantum Gaussian process (QGP) regression was introduced for this task across various classes of unitary evolution. Here, we extend the QGP framework beyond unitary dynamics. In particular, we prove convergence of the channel's outputs to a QGP and derive the associated closed-form kernel under a uniform (Lebesgue measure) prior over quantum channels. 
The kernel's dimensional factor, however, dictates the required observation precision. While manageable when the channel and observable are restricted to small subsystems, exponential suppression precludes learning when the subsystem grows extensively with the system size. Since the Lebesgue prior is overly broad for many applications, we propose an empirical Bayes heuristic that replaces the dimensional factor with a learnable scale parameter while retaining the kernel's state-overlap correlation structure.  
In numerical simulations of up to 64 qubits, channel QGP regression with the Lebesgue kernel exhibits a strong inductive bias for local channels, enabling faithful extrapolation. For global 64-qubit channels, the rescaled kernel restores learnability, with predictions improving systematically with the shot budget.
Results from a noisy quantum computer further demonstrate the robustness of QGP regression under experimental conditions. Beyond regression, we validate QGPs as Bayesian-optimization surrogates for state preparation under noisy XXZ dynamics.
\end{abstract}

\maketitle

\section{Introduction}

Characterizing what a quantum device does to a family of input states is a routine experimental task. In many settings, however, the object of interest is not the full process but a small set of observables measured at the output. Examples include magnetization and correlation functions used to probe quantum many-body dynamics~\cite{dutta2016anti,singh2021driven,kempa2026boundary,lee2026benchmarking}, as well as expectation values that serve as objective functions in variational quantum algorithms and state-preparation protocols~\cite{cerezo2020variationalreview,yoshioka2020variational,chang2025primer}. Similar settings arise in benchmarking, calibration, and characterization experiments, where only selected observables are monitored rather than the full quantum process~\cite{chow2012universal,greenbaum2015introduction,roncallo2023Pauli,nielsen2021gate}. In all these cases, the evolution mapping the inputs to the outputs is most generally an unknown quantum channel. Completely reconstructing this channel by process tomography requires an amount of resources that is exponential in the number of qubits, which is prohibitive and particularly unjustified when the actual goal is to track only a small number of observables.

In this setting, a natural tool for learning such input-output relations is Gaussian process regression, which can predict expectation values for unseen input states from a small training set, while providing native uncertainty quantification. Our goal, however, is not to impose an arbitrary Gaussian process model, but to analytically derive one from the underlying quantum dynamics. This involves specifying a prior mean and covariance function, called a kernel, whose form depends on the ensemble from which the unknown evolution is assumed to be drawn. For unitary dynamics, such physics-based priors are known. Ref.~\cite{garcia2023deep} showed that Pauli expectation values generated by Haar-random global unitaries form Gaussian processes in the large-dimensional limit, and derived the associated kernel exactly. Subsequently, Ref.~\cite{jager2026provable} developed this connection into a Bayesian framework for learning from quantum data and identified analytic kernels for matchgate (free-fermionic) evolutions~\cite{jozsa2008matchgates}, including those acting globally on all qubits, establishing a family of models that are both provable and scalable. These results, however, apply to closed-system dynamics. An experiment or noisy quantum device generally realizes an open-system evolution, that is, a quantum channel, and extending this framework thus requires proving output convergence to Gaussian processes for appropriate priors over channels.

\begin{figure}[t]
    \centering
    \includegraphics[width=1\linewidth]{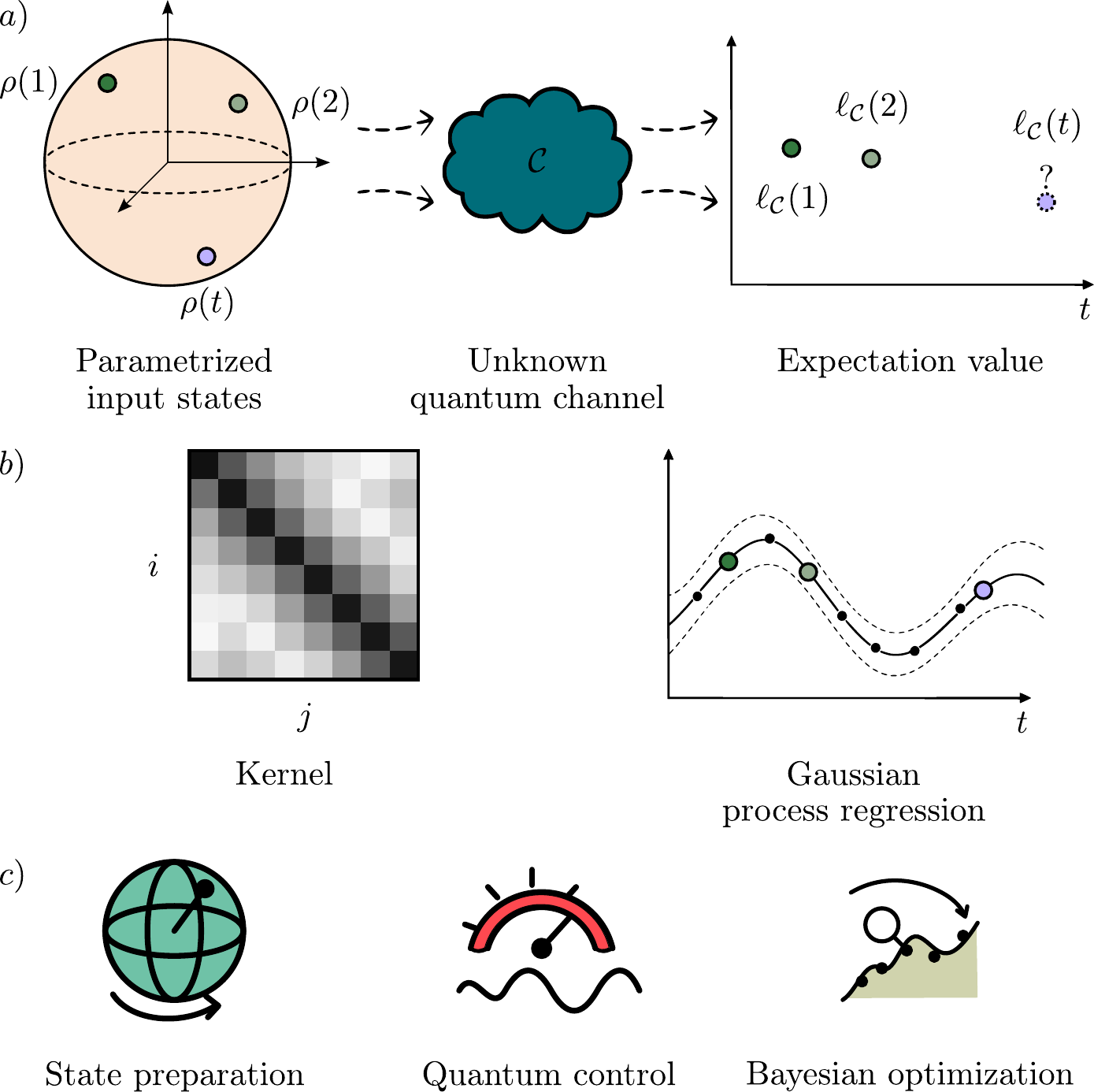}
    \caption{{\bf Quantum Gaussian process regression for quantum channels.}
(a) A parametrized family of input states $\rho(t)$ is acted on by an
unknown quantum channel $\CC$. Expectation values
$\ell_{\CC}(t)=\Tr[\CC(\rho(t))O]$ are measured for a small set of input
states and used to predict the observable for new inputs.
(b) The kernel encodes correlations between the input states. Together
with the measured training data, it defines the QGP posterior used for
regression. In this work, we consider both the channel kernel derived under the assumption of 
uniform channel sampling,  and its rescaled heuristic  counterpart.
(c) Beyond regression, the channel  QGP framework can be applied to tasks such 
as state preparation, quantum control, and Bayesian optimization.}
    \label{fig:schematic} 
\end{figure}

In this work, we prove such convergence and therefore extend the quantum Gaussian process (QGP)~\cite{garcia2023deep,garcia2024architectures,melchor2025quantitative,jager2026provable} construction to quantum channels (see Fig.~\ref{fig:schematic}). Specifically, in the absence of channel-specific knowledge, we take the prior over the unknown evolution to be the Lebesgue measure on the convex set of quantum channels~\cite{kukulski2021generating}. This measure is uniform over channels and provides a probabilistic formulation that avoids favoring any particular region of channel space. Using its equivalent Stinespring representation in terms of Haar-random unitaries on a dilated space, we derive the corresponding QGP prior, which has zero mean and a closed-form kernel, with correlations determined by the pairwise overlaps of the input states. Below, we refer to this kernel as the channel kernel. %

As kernel properties can significantly impact generalization performance and learning efficiency in quantum machine learning \cite{huang2021power, kubler2021inductive}, informing the kernel form by physics-derived insight is of great importance to its applicability. From this perspective, it is noteworthy that although the overlap dependence in the derived kernel is similar to  constructions used in the broader literature on quantum kernel methods~\cite{havlivcek2019supervised, schuld2019quantum}, in the present setting the same overlap structure emerges directly from the assumed prior over quantum channels rather than being postulated as a learning model.

The channel kernel also contains a dimensional prefactor that suppresses its overall scale. This suppression remains manageable when the channel acts only on a sufficiently small subsystem, but becomes exponential when that subsystem grows extensively with the system size. In such a case, the number of measurements required  for successful observable learning grows exponentially with the system size, making applications prohibitively expensive. 
Despite this negative result, experimental applications provide many examples of non-zero  observations at the output of quantum channels acting globally on large systems~\cite{proctor2025benchmarking}. Since, for the Lebesgue prior, the probability of such channel instances decays exponentially with the system size,  this prior lacks inductive bias for predicting observations in such cases. To address this issue, we propose an empirical Bayes heuristic that defines a rescaled kernel, replacing the dimensional prefactor with a trainable hyperparameter while retaining the state-overlap correlation structure of the channel kernel. We optimize this hyperparameter by maximizing the marginal likelihood of the observations, leveraging a closed-form objective native to Gaussian processes (which is unavailable in non-Bayesian approaches), and avoiding any quantum computing overhead. This allows the model to adjust how strongly it relies on similarities between input states when measurements are noisy. In particular, when the channel observations indicate that the correlations are highly informative of the observable values, the predictions from the rescaled kernel exhibit strong inductive bias, irrespective of dimension.  

In our numerical experiments, we first provide a proof-of-principle  demonstration of the  framework with the channel kernel for a family of input states generated by single-qubit parameterized rotations,  4-qubit  channels acting globally (on a 4-qubit system) and locally (on a 64-qubit system), and a Pauli observable.  We implement the channels by joint system-environment Trotterized time evolution with an Ising transverse-field Hamiltonian. We follow with a hardware demonstration of this setup on an IBM quantum computer.  Next, we  numerically compare the performance of the channel and rescaled kernels for 4-qubit and 64-qubit global channels. We find that for the 4-qubit case both perform similarly. In contrast, for the $64$-qubit test case, the rescaled kernel enables successful regression and systematically improves accuracy with increasing shot allocation, while the channel kernel fails to improve upon the prior for the investigated shot budgets. Finally, we demonstrate an application of the framework to Bayesian optimization of a state preparation task.

The remainder of this manuscript is organized as follows. Section~\ref{sec:gp_prelim} reviews the Gaussian process framework needed for
our work. In Section~\ref{sec:channel_qgp}, we derive the Lebesgue QGP prior for quantum
channels (Sec.~\ref{sec:channel_qgp:provable}) and analyze its finite-shot scalability (Sec.~\ref{sec:channel_qgp:scalable}). Section~\ref{sec:channel_GPR_experiments} demonstrates regression with the channel kernel numerically and on quantum hardware. Section~\ref{sec:rescaled_kernel} introduces the rescaled kernel heuristic, and Section~\ref{sec:heuristics} evaluates its application to regression and Bayesian optimization.  We conclude and discuss implications of our results in Section~\ref{sec:discussion}. Additional proofs, implementation details, and numerical results are provided in the appendices.

\section{Gaussian Process Preliminaries}
\label{sec:gp_prelim}

We begin by recalling the basic Gaussian process formalism used throughout
this work. Let $f=\{f(t)\}_{t\in\mathscr{I}}$ be a stochastic process
indexed by $t\in\mathscr{I}$. We say that $f$ forms a Gaussian process
(GP), denoted by
\begin{equation}
    f\sim\mathcal{GP}(\mu,\kappa),
\end{equation}
if and only if, for any finite set
$T=\{t_1,\ldots,t_k\}\subseteq\mathscr{I}$, the random vector
\begin{equation}
    \vec{f}=
    \left(
        f(t_1),\ldots,f(t_k)
    \right)^\top
\end{equation}
follows a multivariate Gaussian distribution
\begin{equation}
    \vec{f}\sim
    \NC(\vec{\mu},\Sigma).
    \label{eq:GP}
\end{equation}
Here, the entries of the mean vector and covariance matrix are
\begin{equation}
    \mu_i=\mu(t_i),
    \qquad
    \Sigma_{ij}=\kappa(t_i,t_j)
    =
    {\rm Cov}\left[f(t_i),f(t_j)\right].
    \label{eq:GP_mean_cov}
\end{equation}
The pair $(\mu,\kappa)$ defines the GP prior.

In practice, the training labels are obtained with finite precision. We
therefore consider observations
\begin{equation}
    \vec{y}
    =
    \vec{f}+\vec{\epsilon},
    \qquad
    \vec{\epsilon}\sim\NC(\vec{0},\Gamma),
    \label{eq:GP_observations}
\end{equation}
where $\Gamma$ denotes the noise covariance matrix. For
independent observations, $\Gamma$ is diagonal. In the setting considered
below, its entries account for the finite-shot uncertainty in estimating
expectation values.

Consider now a new point $t\notin T$, and define the covariance vector
$\vec{m}(t)$ with entries
\begin{equation}
    m_i(t)=\kappa(t,t_i).
\end{equation}
The joint distribution of $f(t)$ and the observations is
\begin{equation}
    \begin{pmatrix}
        f(t)\\
        \vec{y}
    \end{pmatrix}
    \sim
    \NC\left[
    \begin{pmatrix}
        \mu(t)\\
        \vec{\mu}
    \end{pmatrix},
    \begin{pmatrix}
        \kappa(t,t) & \vec{m}(t)^\top\\
        \vec{m}(t) & \Sigma+\Gamma
    \end{pmatrix}
    \right].
    \label{eq:GP_joint}
\end{equation}
Conditioning on the observations gives the posterior distribution
\begin{equation}
    p\left(f(t)\mid\vec{y}\right)
    =
    \NC\left(\mu_P(t),\sigma_P^2(t)\right),
    \label{eq:GP_posterior}
\end{equation}
with
\begin{align}
    \mu_P(t)
    &=
    \mu(t)
    +
    \vec{m}(t)^\top
    \left(\Sigma+\Gamma\right)^{-1}
    \left(\vec{y}-\vec{\mu}\right),
    \label{eq:normal-kp1}
    \\
    \sigma_P^2(t)
    &=
    \kappa(t,t)
    -
    \vec{m}(t)^\top
    \left(\Sigma+\Gamma\right)^{-1}
    \vec{m}(t).
    \label{eq:normal-kp2}
\end{align}
The posterior mean provides a prediction for $f(t)$, while the posterior
variance quantifies the uncertainty associated with that prediction.
The latter can also be used to select new points at which information
should be acquired, as in active learning and Bayesian optimization (BO)~\cite{jones1998efficient,snoek2012practical}.

\section{Quantum Gaussian Processes for Quantum Channels}
\label{sec:channel_qgp}

We now turn to the quantum learning problem considered in this work.
Given a family of input states $\{\rho(t)\}_{t\in\mathscr{I}}$, an underlying
unknown quantum channel $\CC$, and an observable $O$, each input state
is associated with the expectation value
$\ell_{\CC}(t)=\Tr[\CC(\rho(t))O]$. If the unknown channel is regarded
as being drawn from an ensemble of possible quantum evolutions, then
$\{\ell_{\CC}(t)\}_{t\in\mathscr{I}}$ defines a stochastic process.
Whenever this stochastic process forms a Gaussian process, its prior
mean and covariance may be determined from the underlying ensemble of
channels, specifying a physics-informed model that we refer to as a provable
quantum GP.

To begin, let us define
\begin{equation}
    \HC_A=(\mathbb{C}^2)^{\otimes n_A},
    \qquad
    \HC_B=(\mathbb{C}^2)^{\otimes n_B},
\end{equation}
and consider a quantum channel
\begin{equation}
    \CC:\BC(\HC_A)\rightarrow\BC(\HC_B),
\end{equation}
where $\BC(\HC)$ denotes the set of bounded linear operators acting on $\HC$.
We take a family of input states
\begin{equation}
    \SC=\{\rho(t)\}_{t\in\mathscr{I}},
\end{equation}
indexed by $t\in\mathscr{I}$\footnote{While it is intuitive to think of $t$ as time, it can represent an arbitrary, potentially multidimensional parameter with no direct physical interpretation.}, and a Pauli observable
$O\in\BC(\HC_B)$. For each input state, we define
\begin{equation}
    \ell_{\CC}(t)
    =
    \Tr\left[\CC(\rho(t))O\right].
    \label{eq:exp}
\end{equation}
More generally, $O$ can be any traceless Hermitian operator satisfying
$O^2\propto\id$.

The basic regression task considered in this work is to predict
$\ell_{\CC}(t)$ from a finite dataset
\begin{equation}
    \DC_k
    =
    \left\{
        \left(\rho(t_i),y_i\right)
    \right\}_{i=1}^k,
    \label{eq:training_dataset}
\end{equation}
where $y_i$ is a finite-shot estimate of $\ell_{\CC}(t_i)$. The
observations form the vector $\vec{y}$ introduced in
Eq.~\eqref{eq:GP_observations}, while $\Gamma$ contains their
finite-shot measurement uncertainties.

\subsection{Provable channel QGPs}
\label{sec:channel_qgp:provable}

As previously mentioned, to construct a QGP, we regard the unknown channel $\CC$ as being drawn
from a set of channels $\CS$ according to a probability measure $\nu$, i.e., as a stochastic process. If this process forms a GP, the corresponding
QGP prior is specified by
\begin{align}
    \mu(t)
    &=
    \mathbb{E}_{\CC\sim\nu}
    \left[
        \ell_{\CC}(t)
    \right],
    \label{eq:qgp_mean}
    \\
    \kappa(t_i,t_j)
    &=
    {\rm Cov}_{\CC\sim\nu}
    \left[
        \ell_{\CC}(t_i),
        \ell_{\CC}(t_j)
    \right].
    \label{eq:qgp_kernel}
\end{align}
Thus, constructing the QGP amounts to determining the prior mean and
kernel associated with the assumed channel ensemble and measure. Once these are
known, QGP predictions follow directly from
Eqs.~\eqref{eq:normal-kp1}--\eqref{eq:normal-kp2}.

When $\CC$ is a unitary channel,
$\CC(\rho)=U\rho U^\dagger$, this setting reduces to that studied in
Refs.~\cite{garcia2023deep,jager2026provable}. Here, we instead consider
general quantum channels. The first step in constructing the corresponding
QGP is therefore to choose a probability measure $\nu$ over the set
$\CS$ of possible channels. In the absence of channel-specific
knowledge, we take $\nu$ to be the Lebesgue measure over the convex set
of quantum channels. This corresponds to sampling uniformly from the set
of all channels and, in this sense, does not privilege any particular
channel a priori~\cite{kukulski2021generating}.

The Lebesgue measure admits a convenient representation through the
Stinespring dilation~\cite{kukulski2021generating}. Let
\begin{equation}
    \HC_F=(\mathbb{C}^2)^{\otimes 2n_B},
    \qquad
    \HC_E=(\mathbb{C}^2)^{\otimes(n_A+n_B)},
\end{equation}
be auxiliary input and output Hilbert spaces, respectively, and define
\begin{equation}
    d=2^{n_A+2n_B}.
\end{equation}
Then, for a fixed pure state $\dyad{\io}_F$, the expectation value in
Eq.~\eqref{eq:exp} can be expressed as
\begin{equation}
    \ell_{\CC}(t)
    =
    \Tr\left[
        \left(O\otimes\id_E\right)
        U_{\CC}
        \left(\rho(t)\otimes\dyad{\io}_F\right)
        U_{\CC}^{\dagger}
    \right],
    \label{eq:exp_stinespring}
\end{equation}
where $U_{\CC}$ is a unitary mapping
$\HC_A\otimes\HC_F$ to $\HC_B\otimes\HC_E$. Sampling $\CC$ according
to the Lebesgue measure is equivalent to sampling $U_{\CC}$ according
to the Haar measure over $\mathbb{U}(d)$ and tracing out $\HC_E$~\cite{kukulski2021generating}. Hence, the averages defining
the QGP prior in Eqs.~\eqref{eq:qgp_mean} and~\eqref{eq:qgp_kernel}
can be evaluated using Haar integration and Weingarten
calculus~\cite{collins2006integration,mele2023introduction}.

In particular, using the asymptotic Haar-moment results of
Ref.~\cite{garcia2023deep} in the limit of large system size (i.e., $n_A+2n_B \to \infty$), we find that for a Pauli observable $O$, the
QGP prior has zero mean,
\begin{equation}
    \mu(t)
    =
    \mathbb{E}_{\CC\sim\nu}
    \left[
        \ell_{\CC}(t)
    \right]
    =
    0,
    \label{eq:Lebesgue_mu}
\end{equation}
and a kernel given by
\begin{equation}
    \kappa(t_i,t_j)
    =
     \mathbb{E}_{\CC\sim\nu}
    \left[
        \ell_{\CC}(t_i)
        \ell_{\CC}(t_j)
    \right]
    =
    \frac{1}{d}
        \Tr[\rho(t_i)\rho(t_j)].
    \label{eq:Lebesgue_kernel}
\end{equation}
Thus, the kernel (covariance) matrix entering the QGP posterior in Eq.~\eqref{eq:GP_posterior} has entries
$\Sigma_{ij}=\kappa(t_i,t_j)$. Equation~\eqref{eq:Lebesgue_kernel}
provides a closed-form and hyperparameter-free kernel whose correlations
are completely determined by the pairwise overlaps of the input states. 
The form of Eq.~\eqref{eq:Lebesgue_kernel} also makes the kernel
experimentally accessible. Its entries require estimating overlaps
$\Tr[\rho(t_i)\rho(t_j)]$, which can be obtained, for instance, using a
SWAP test~\cite{barenco1997stabilization,buhrman2001quantum,cincio2018learning}. A standard
implementation acting on $n_A$-qubit states requires $2n_A+1$ qubits and
a circuit of depth $\mathcal{O}(n_A)$. Thus, constructing the kernel does
not require learning the channel itself, but only estimating pairwise
similarities of the input states.

Importantly, we recall that the results of Ref.~\cite{garcia2023deep} guarantee convergence to a QGP  with the kernel given in Eq.~\eqref{eq:Lebesgue_kernel} provided that the input states satisfy specific conditions. In particular, we find a QGP with positive correlations when $ \Tr[\rho(t_i)\rho(t_j)]\in\Omega(1/\poly(n_A,n_B))$. 
We also stress that the use of the Lebesgue measure does not require the
physical channel implemented by a device to have been generated at
random. Rather, it specifies the probabilistic uniform prior used when no more detailed
information about the channel is available. The predictions for the outputs of the
implemented channel  are then inferred from the observations in the training set
through the QGP posterior.

\subsection{Scalable subsystem channel QGPs}
\label{sec:channel_qgp:scalable}

The number of measurement shots required for effective learning is dictated by the magnitude of the noise in the observations (arising from finite sampling) relative to the kernel scale. For a Pauli observation from $N_i$ measurements, the diagonal entry of the noise matrix satisfies
\begin{equation}
    \Gamma_{ii}
    =
    {\rm Var}\!\left[y_i\right]
    =
    \frac{1-\ell_{\CC}(t_i)^2}{N_i}
    \leq
    \frac{1}{N_i}.
    \label{eq:shot_noise_scaling}
\end{equation}

Because the channel kernel in Eq.~\eqref{eq:Lebesgue_kernel} scales with an explicit dimensional factor of $1/d$, we can isolate this prefactor from the state-dependent correlations, which match the overlap (or fidelity) kernel for input states \cite{havlivcek2019supervised, schuld2019quantum},
\begin{equation}
    \kappa_{\rm ov}(t_i, t_j)
    =
    \Tr[\rho(t_i)\rho(t_j)].
    \label{eq:overlap_kernel}
\end{equation}
Substituting $\Sigma = \frac{1}{d}\Sigma^{\rm ov}$ and $\vec{m}(t) = \frac{1}{d}\vec{m}^{\rm ov}(t)$ into the QGP posterior, specifically Eqs.~\eqref{eq:normal-kp1}--\eqref{eq:normal-kp2}, explicitly reveals how this $1/d$ scaling impacts the predictions:
\begin{equation}
    \mu_{P}(t)
    =
        \vec{m}^{\rm ov}(t)^\top
    \left(
        \Sigma^{\rm ov}
         +
        d{\Gamma}
    \right)^{-1}
    \vec{y} .
    \label{eq:Leb_posterior_rescaled}
\end{equation}
Here, the prior $1/d$ scaling factor effectively amplifies the observation noise matrix to $\Gamma_{\rm eff} = d\Gamma$ \footnote{The variance $
    \sigma_{P}^{2}(t)
    =
    \frac{1}{d}
    [ 
    \Tr[\rho(t)^2] 
    -
    \vec{m}^{\rm ov}(t)^\top
    (
        \Sigma^{\rm ov}
          +  
        d{\Gamma}
    )^{-1}
    \vec{m}^{\rm ov}(t)
    ]
$ reveals the same effective observation noise amplification.}.
Appendix~\ref{app:finite_shot_stability} further analyzes the stability of the posterior in the presence of this noise, including noisy kernel estimates. 

This effective noise amplification remains efficiently manageable only when the relevant channel effectively acts on a sufficiently small subsystem, that is, when the observable expectation value  is affected by small subsystems of the output and  the input state. 
If the target expectation value depends solely on a channel mapping an input subsystem $A$ (of size $n_A$) to an output subsystem $B$ (of size $n_B$), the relevant input states are the reduced density matrices $\rho_A(t) = \Tr_{\bar{A}}[\rho(t)]$. Accordingly, the subsystem channel kernel is given by the overlap of the reduced input states,
\begin{equation}
\label{eq:Lebesgue_kernel_subsystem}
\kappa_{\rm sub}(t_i,t_j)=\frac{1}{d}\operatorname{Tr}\left[\rho_A(t_i)\rho_A(t_j)\right].
\end{equation}
Since it depends only on reduced-state representations of the inputs, this kernel can be viewed as a linear projected quantum kernel~\cite{huang2021power} with a $1/d$ prefactor. The dimension factor $d=2^{n_A+2n_B}$ now depends only on the input and output subsystem sizes.
If this subsystem grows at most logarithmically with the total system size $n$, i.e.,
\begin{equation}
n_A + 2n_B \in \mathcal{O}(\log n),
\end{equation}
 $d$ remains polynomial. Polynomial shot resources then suffice to bound the effective observation noise $\Gamma_{\rm eff} \in \OC(1/\poly(n))$ and to resolve the kernel against the noise.

Conversely, if the  channel effectively involves an extensive fraction of the system, e.g., $n_A + 2n_B \in \Theta(n)$, the dimension $d$ grows exponentially. An exponentially growing shot budget is then required to dampen $\Gamma_{\rm eff}$ strongly enough. Otherwise, $\Gamma_{\rm eff}$ asymptotically dominates the state-overlap signal $\Sigma^{\rm ov}$, pushing predictions toward the uninformative zero-mean prior. A heuristic approach in Sec.~\ref{sec:rescaled_kernel} uses an empirical Bayes method to restore learnability in such extensive systems for favorable datasets.

\section{Channel QGP regression experiments}
\label{sec:channel_GPR_experiments}

In this section, we test regression using the channel QGP derived from the
Lebesgue prior. 
We hence verify that finite-shot QGP inference succeeds in the favorable regime. Importantly, this experimental validation is not only based on numerical simulations (Sec.~\ref{sec:channel_GPR_numerics}) but also includes a real-device implementation (Sec.~\ref{sec:channel_GPR_IBM}).
More precisely, as we work here with finite $n_A$ systems, we
employ a non-asymptotic kernel variant 
\begin{equation}
    \kappa(t_i,t_j)    
    = 
    \frac{d}{d^2-1}
    \left(
        \Tr[\rho_A(t_i)\rho_A(t_j)]
        -
        \frac{1}{d} 
    \right),    
    \label{eq:Lebesgue_kernel_finite_n}
\end{equation}
derived in Ref.~\cite{garcia2023deep}, which for large subsystems  converges to the asymptotic form of Eq.~\eqref{eq:Lebesgue_kernel}.
Appendix~\ref{app:GPR_impl} details our QGP  regression implementation.

\subsection{Numerical results for regression}
\label{sec:channel_GPR_numerics}

We consider a small ($n=4$ qubits) and large ($n=64$ qubits) system with a family of pure input states $\rho(t) = \dyad{\psi(t)}$, where
\begin{equation}
\ket{\psi(t)}
=
U_S
\bigg(
\bigotimes_{j=0}^{n-1}
e^{-i\phi(j,t)Y_j}
e^{-i\phi(j,t)X_j}
\ket{0}_j
\bigg)
\label{eq:Ising_input}
\end{equation}
is generated by site- and parameter-dependent rotations around the $X$ and
$Y$ axes, followed by parameter-independent system unitary dynamics $U_S$.
The latter is generated by a one-dimensional transverse-field Ising Hamiltonian using a
Trotter product formula.
To implement a channel, the input states are subsequently coupled to an environment through
joint system-environment unitary dynamics. For $n=4$, all system qubits are
coupled to an environment with $n_{\rm env}=4$ qubits. For $n=64$, only
four neighboring system qubits are coupled to a four-qubit environment. In both
cases, the coupling is implemented by Trotterized transverse-field one-dimensional Ising
time evolution. The learning targets are the Pauli expectation values $\langle X_1\rangle(t)$ for the $n=4$ system
and $\langle X_{60}\rangle(t)$ for the $n=64$ system. The first
case describes a global channel acting on the full, albeit small, system, while the
second tests the subsystem regime in which a large quantum system is  acted upon by a low-dimensional channel. Further details are provided
in Appendix~\ref{app:Ising_impl}.

\begin{figure*}[t!]
\includegraphics[width=.99\linewidth]{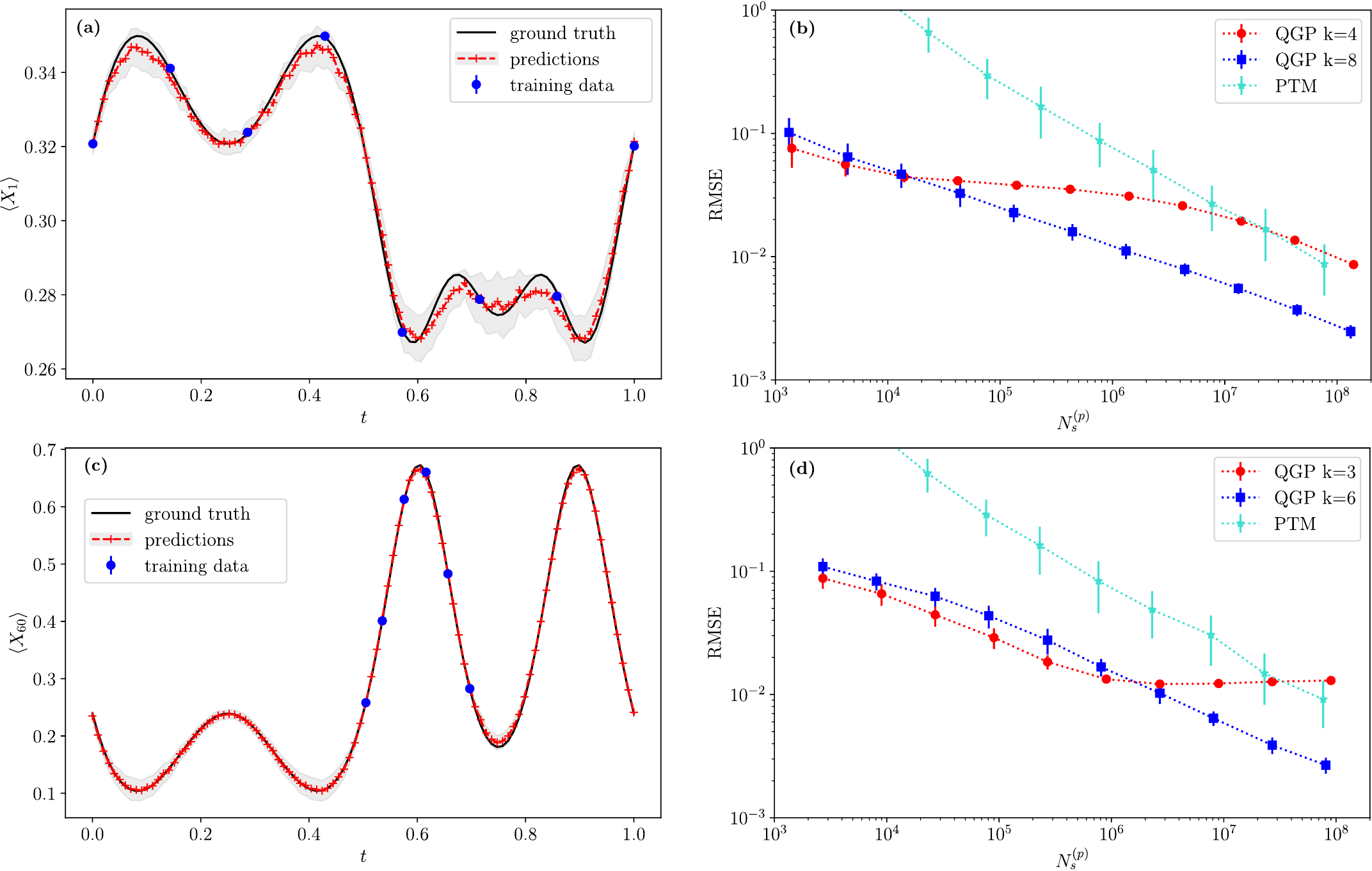}
\caption{{\bf QGP regression with the  channel kernel.}
We learn the expectation value of $X_1$ ($X_{60}$) for a family of
4-qubit (64-qubit) input states generated by single-qubit rotations
parametrized by $t$ followed by 
system unitary dynamics. Joint system-environment 
time evolution implements the channel. 
Both are implemented by a transverse-field one-dimensional Ising Hamiltonian and a Trotter product formula.  
For $n=4$, the channel acts globally on all system qubits,
while for $n=64$ it acts locally on a 4-qubit subsystem.
In (a), we show QGP regression for the global channel with the shot budget $N_s^{(p)}=1.32\cdot10^8$ for a single
prediction. The black curve is the ground
truth, the blue circles are the training data (including error bars representing their finite-shot uncertainty), the red pluses are the
predictions, and the shaded region denotes the $95\%$ prediction confidence interval, with a half-width $2 \sigma_P(t)$.
In (b), we show the corresponding root mean squared error (RMSE) versus
$N_s^{(p)}$ for $k=4$ and $k=8$ training observations, together with the RMSE
of row-PTM learning as a reference. In (c) and (d), we show the analogous
results for the local channel, with
$N_s^{(p)}=2.7\cdot10^7$ in (c) and $k\in\lbrace 3, 6 \rbrace$ in (d).
To account for finite-shot fluctuations, all experiments are evaluated across 100 independent runs. Hence, (a,c) display regression instances with the median RMSE, while (b,d) show the mean RMSE (the markers) and sample standard deviation (the error bars).}
\label{fig:GPR_Haar}
\end{figure*}

For $n=4$, we consider an interpolation task, while for $n=64$ we
consider the more challenging problem of extrapolation. For interpolation
we use $k\in\{4,8\}$ training observations with $t$ on a uniform grid over
$[0,1]$. For extrapolation we use datasets  with  $k\in\{3,6\}$ and $T$ forming a
uniform grid on a narrow region of $[0.5,0.7]$. In both cases, predictions are made over
the full interval $t\in[0,1]$.
For both learning problems, the channel has $n_A=4$, while the
observable acts on a single output qubit, and hence $n_B=1$. We therefore
use the channel kernel~(\ref{eq:Lebesgue_kernel_finite_n}) with
$d=64$.

Since the dominant resource in practice is the number of
measurements, we study the dependence on the shot budget in detail. We
consider a shot budget with
$
    1320\leq N_s^{(p)}\leq1.4\cdot10^8
$
shots necessary for a single prediction. We use finite shot estimates of the observations $\vec{y}$, non-diagonal part of the covariance matrix $\Sigma$, and the covariance
vector $\vec{m}$. We note that $\vec{y}$ and $\Sigma$ can be reused for multiple predictions, while $\vec{m}$ needs to be estimated independently for each prediction.  Here, each independent covariance function estimate and each training observation use the same shot number. Thus, the total experiment shot cost is  $\frac{k+1+2k_p}{k+3}N_s^{(p)}$, where $k_p$ is the number of observable predictions. 
We quantify the prediction accuracy over a uniform grid of $k_p=100$ values
of $t$ on $[0,1]$ using the root mean squared error (RMSE).
Because finite-shot sampling makes the RMSE a random variable, we report its mean and standard deviation across 100 independent experiment runs.  

Figure~\ref{fig:GPR_Haar}(a,c) shows that both $4$-qubit global channel interpolation ($k=8, N_s^{(p)}=1.32\cdot10^8$) and $64$-qubit local channel extrapolation ($k=6, N_s^{(p)}=2.1\cdot10^7$) yield accurate predictions. The latter highlights the impact of the subsystem inductive bias, which successfully enables the QGP to extrapolate $\langle X_{60}\rangle(t)$ over the full prediction interval.
In both cases, lower shot budgets already recover the dominant features of the
observable dependence, as shown in Appendix~\ref{app:GPR_low_shots}. 
Figure~\ref{fig:GPR_Haar}(b,d) summarizes the
systematic dependence of the RMSE on $N_s^{(p)}$ and $k$. For all values of $k$, the RMSE decreases
with the shot budget for $N_s^{(p)}<10^6$. For the $n=64$ system, the
error plateaus at larger shot budgets for $k=3$, whereas for $k=6$ it
continues to decrease. In this case, the accuracy is therefore limited
by the size of the training data rather than by the available shots. For
the four-qubit global channel, both $k=4$ and $k=8$ continue to improve
over the full range of shot budgets considered, with the larger training
set giving consistently smaller errors at high shot counts. 

We also benchmarked the channel QGP against a row Pauli transfer matrix (PTM) method, which learns
the row of the Pauli transfer matrix associated with the observable~\cite{nielsen2021gate,roncallo2023Pauli}. To obtain the  prediction, this
row is applied to the Pauli representation of $\rho_A(t)$,
which is obtained by state tomography. This reference approach is described
in Appendix~\ref{app:PTM}. We evaluate the row-PTM approach using the same
values of $t$ and comparable shots required for a single prediction
$N_s^{(p)}$, and also report its RMSE in
Fig.~\ref{fig:GPR_Haar}(b,d). For the larger training sets, QGP regression
achieves smaller RMSE in both systems and across all the shot budgets considered. Reaching a given
RMSE with the QGP regression typically requires one to two orders of magnitude fewer shots than with
the row-PTM approach.

\subsection{Real-device implementation}
\label{sec:channel_GPR_IBM}

Moving beyond numerical simulations, we implement a 4-qubit global channel   on IBM's \texttt{ibm\_boston} quantum computer. The channel is generated by a quantum circuit implementing Trotterized transverse-field Ising system-environment dynamics which couples the system to a 4-qubit environment, similarly to the numerical experiments above. 
The implementation is inevitably affected by  unknown hardware noise, which does not pose a problem to our QGP approach but is instead naturally accommodated, since the prior does not assume a particular channel form.
We use input states prepared by circuits implementing a parametrized family of states from  Eq.~(\ref{eq:Ising_input})  with $U_S=\id$.   Further implementation details are available in Appendix~\ref{app:Ising_impl}. 

\begin{figure}[t]
\includegraphics[width=.99\linewidth]{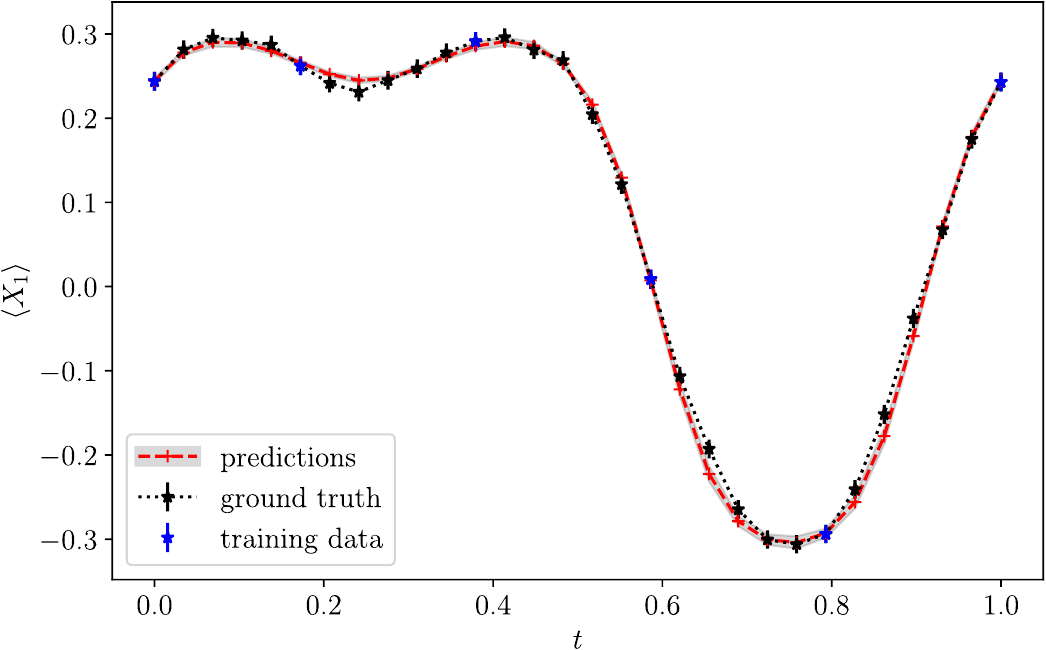}
\caption{{\bf QGP regression with the channel kernel on IBM's
quantum computer \texttt{ibm\_boston}.}
We predict $\langle X_1\rangle(t)$ for a family
of 4-qubit input states generated by single-qubit rotations parametrized
by $t$.  The channel
is a real-device implementation of a circuit performing Trotterized transverse-field Ising dynamics  coupling the system to a 4-qubit environment. The ground truth, indicated by the black markers, are the results obtained from
the quantum device with 34400 shots per data point.  The error bars, computed as twice the standard deviation,  quantify the finite shot effects. The 
blue markers are the training data obtained from the same device and with the same number of shots.  
 The red pluses are the
QGP predictions and the shaded region denotes the 95\% prediction confidence interval. The kernel is evaluated numerically without shot noise effects, neglecting the hardware noise  on the input states.}
\label{fig:channel_GPR_IBM}
\end{figure}

We demonstrate  channel QGP  regression interpolating  $\langle X_1\rangle (t)$ on the interval $t \in [0, 1]$. We emphasize that we interpolate 
the real-hardware outputs rather than the ideal channel outputs unaffected by the hardware noise. We use 6 training observations with values of $t$ distributed on a uniform grid covering the $[0,1]$ interval. Each observation is estimated with 34400 shots. We approximate the kernel entries by overlaps computed numerically using an exact state simulator while neglecting the hardware noise. This approximation is justified because the state-preparation circuits only contain single-qubit gates and hence are not expected to introduce significant errors, while estimating the overlaps via SWAP-test circuits on actual hardware would likely be too noisy and preclude learning. We test the interpolation by predicting $\langle X_1 \rangle$ for $19$ values of $t$ uniformly spaced between 0 and 1, and compare them to the estimates obtained from the device with $34400$ shots per $t$ value. The results presented in Fig.~\ref{fig:channel_GPR_IBM} show that the predictions agree well with the experimental outcomes and accurately reproduce the parameter-observable dependence.

\section{Learning the scale of the channel kernel}
\label{sec:rescaled_kernel}

As established in Sec.~\ref{sec:channel_qgp:scalable}, the Lebesgue prior lacks the inductive bias needed to predict observables for high-dimensional quantum channels acting on an extensive fraction of the system. 
Crucially, this is a finite-shot limitation of the prior's rigid overall scale, not a failure of the prior's state-overlap correlations themselves. 
The prior scale implies that  the observations concentrate exponentially around 0 with the system qubit count, as per Eqs.~\eqref{eq:Lebesgue_mu}--\eqref{eq:Lebesgue_kernel}. Nevertheless, in applications we frequently encounter channels that have measurably non-zero observations. To learn such observations with a QGP, we need to modify the prior.

As the state-overlap correlations physically quantify state distinguishability, we retain this correlation structure and instead propose to  empirically tune the dimensional prefactor.    Thus, we introduce a heuristic rescaled channel kernel that replaces $d$ with a learnable scale parameter $\eta$:
\begin{equation}
    \kappa^{\eta}(t_i, t_j) = \eta \Tr[\rho(t_i)\rho(t_j)],
    \label{eq:rescaled_channel_kernel}
\end{equation}
yielding the rescaled training kernel matrix $\Sigma^{\eta} = \eta \Sigma^{\rm ov}$. When $\eta$ is large enough, the rescaled kernel can restore learning feasibility for favorable data sets and high-dimensional channels.

We determine $\eta$ via the empirical Bayes approach of maximizing the log-marginal likelihood $\log p(\vec{y}|T,\eta)$ over $\eta\in[1/d, 1]$,
which admits a closed-form expression \cite{rasmussen2006gaussian} in the Gaussian process case as
\begin{equation}
    \begin{split}
        \log p(\vec{y}|T,\eta) &= -\tfrac{1}{2} \vec{y}^\top \left(\Sigma^{\eta} + \Gamma\right)^{-1} \vec{y} \\
        &\quad - \tfrac{1}{2} \log |\Sigma^{\eta} + \Gamma| - \tfrac{k}{2} \log(2\pi),
    \end{split}
    \label{eq:marginal_likelihood}
\end{equation}
where $|\cdot|$ is the determinant and $k$ is the  training data size.  
The marginal likelihood evaluates the probability of the observations integrated over the entire prior channel ensemble, i.e., 
\begin{equation}
    p(\vec{y} | T, \eta) = \int p(\vec{y} | \vec{f}, T) p(\vec{f} | T, \eta) \, d\vec{f}.
\end{equation}
Consequently, maximizing this objective identifies the channel prior scale $\eta$, rather than a specific channel, under which the observed expectation values are most plausible.

Note that once the $k \times k$ state-overlap matrix $\Sigma^{\rm ov}$ is estimated, maximizing Eq.~\eqref{eq:marginal_likelihood} is a strictly classical routine. Therefore, for favorable datasets, where the input state overlaps are large relative to the shot noise, this procedure yields a data-driven scale parameter without incurring any additional quantum computing overhead.

\section{Applications with the rescaled channel kernel}
\label{sec:heuristics}

Returning to the experimental validation of channel QGPs, we demonstrate how learning the kernel scale enables accurate regression for global channels in large systems (Sec.~\ref{sec:heur_GPR_numerics}), which is otherwise intractable with the  channel kernel. Beyond regression, we then deploy these QGPs as surrogate models for Bayesian optimization (Sec.~\ref{sec:bo_vs_gd}).

\subsection{Numerical results for regression}
\label{sec:heur_GPR_numerics}

\begin{figure*}[t]
\includegraphics[width=.99\linewidth]{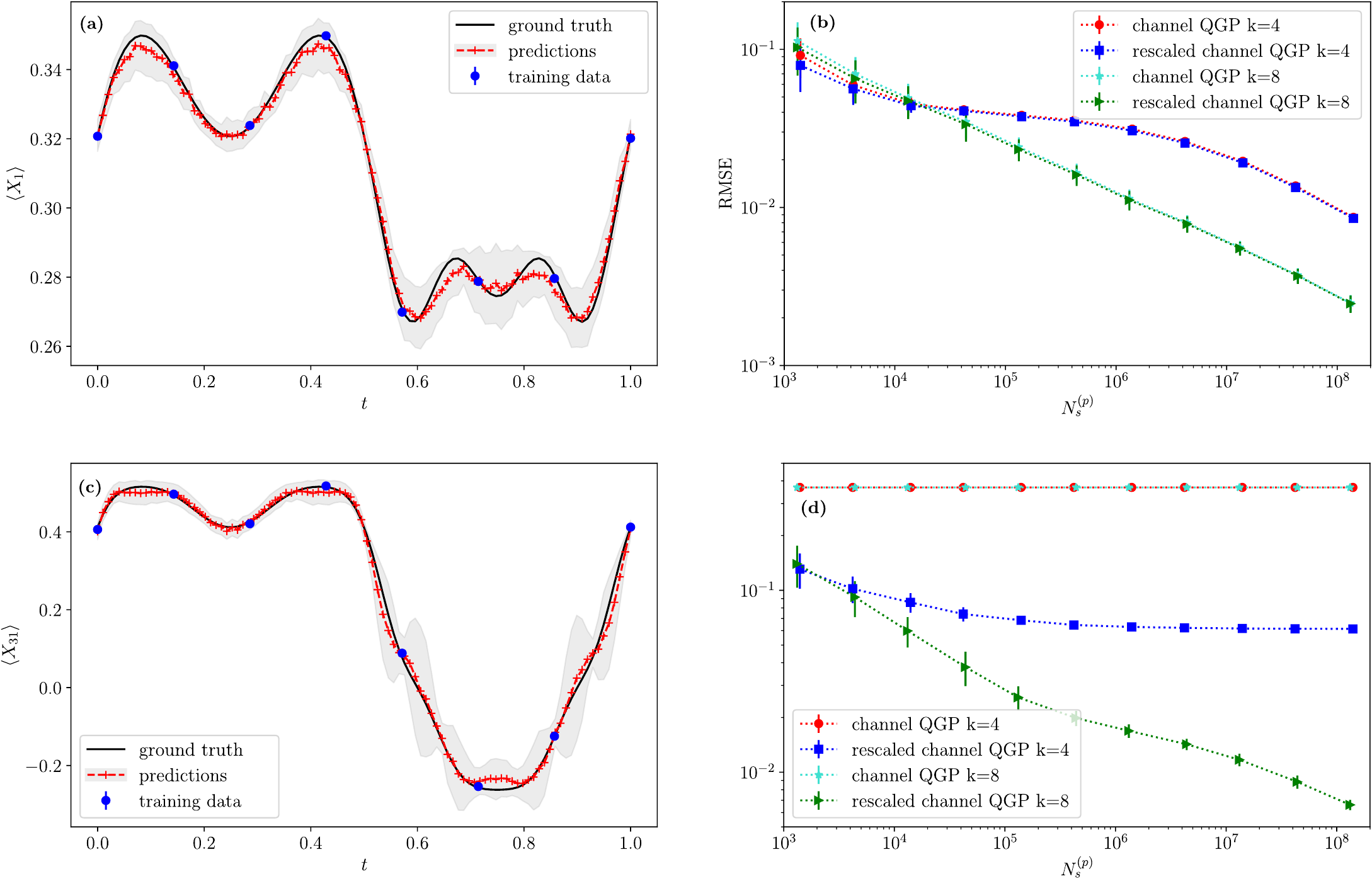}
\caption{ { \bf  QGP regression with the  rescaled channel  kernel}. In (a), we show learning of $\langle X_1 \rangle$  for the same 4-qubit setup as in Fig.~\ref{fig:GPR_Haar}, with shot budget  $N_s^{(p)}=1.32\cdot10^8$ per prediction. In (c), $\langle X_{31} \rangle$ was learned for a 64-qubit family of input states prepared by single-qubit rotations parametrized by $t$, and a global channel generated by one-dimensional Trotterized transverse-field Ising dynamics coupling the system to a 64-qubit environment.  Here, $N_s^{(p)}=1.32\cdot10^6$.  In (a) and (c), we plot the ground truth (the black lines), the training  data (the blue circles), the predictions (the red crosses)  and their $95\%$ confidence intervals (the shaded areas).  In (b) and (d), we compare the RMSE of the rescaled channel  QGP learning and the channel QGP learning for $n=4$ and $n=64$, respectively. For both methods, we plot $k=4$ and $k=8$ RMSE values versus $N_s^{(p)}$.
For $n=4$, the channel QGP (the red circles and the cyan asterisks) gives very similar results to 
the rescaled channel QGP (the blue squares and the green triangles), as shown by the overlapping markers.
For $n=64$, the channel QGP remains at approximately the same RMSE
for both training-set sizes and does not improve as the shot budget is
increased. In contrast, the rescaled channel QGP reproduces the
observable dependence across the prediction interval and its RMSE
decreases systematically with increasing shot budget. To account for the shot noise effects, in (a) and (c) we show learning instances with median RMSE  chosen from a sample of $100$ learning instances. The error bars of the training data, the RMSE estimates and the prediction confidence intervals  are computed as in Fig.~\ref{fig:GPR_Haar}. We also note that reference channel QGP RMSE data for $n=4$ are taken from Fig.~\ref{fig:GPR_Haar}.}
\label{fig:rescaled_channel_GPR}
\end{figure*}

We first apply the rescaled channel kernel to the 4-qubit learning task
from Section~\ref{sec:channel_GPR_numerics}. In contrast to the 
channel kernel considered in the experiments presented before, here the 
kernel scale is determined from the training data, while its dependence
on the input-state overlaps is fixed according to
Eq.~\eqref{eq:rescaled_channel_kernel}. We find that the resulting QGP
learns the observable dependence on $t$ with very similar accuracy
to the channel QGP. An example for $k=8$ and
$N_s^{(p)}=1.32\cdot10^8$ is shown in
Fig.~\ref{fig:rescaled_channel_GPR}(a), while
Fig.~\ref{fig:rescaled_channel_GPR}(b) compares the RMSE obtained with
the two kernels over the full range of shot budgets.

We next test the regime in which the dimensional suppression of the
channel kernel becomes prohibitive. We scale both the system and
environment to $64$ qubits and learn $\langle X_{31}\rangle(t)$ for the
$t$-parametrized family of input states in
Eq.~\eqref{eq:Ising_input}. In this case, $n_A=64$, such that the 
channel itself grows extensively with the system size (see Appendix~\ref{app:Ising_impl}).
We emphasize that, for $n_A=64$, standard process tomography and
Pauli transfer matrix learning are already prohibitively costly. 
We train on $k=4$ and $k=8$ uniformly spaced points on $t \in [0,1]$ (including the interval's endpoints), and quantify the prediction accuracy via the RMSE over 100 test points spaced uniformly between 0 and 1.  

For this $n=64$ setup, Fig.~\ref{fig:rescaled_channel_GPR}(c) shows
successful learning with $k=8$ training points and
$N_s^{(p)}=1.32\cdot10^6$. The drastically different finite-shot behavior of the
two kernels is evident in Fig.~\ref{fig:rescaled_channel_GPR}(d). The
standard channel QGP fails, plateauing at ${\rm RMSE}\approx0.368$ across both datasets and over the full
range $1320\leq N_s^{(p)}\leq1.4\cdot10^8$. In contrast, the error obtained
with the rescaled kernel decreases systematically with the shot budget.
For $k=4$, the RMSE reaches $0.06$ for $N_s^{(p)}>10^5$. Increasing
the training set to $k=8$ further reduces the error to
$5\cdot10^{-3}$ at $N_s^{(p)}=1.32\cdot10^8$.

\subsection{Bayesian optimization for noisy state preparation}
\label{sec:bo_vs_gd}

Gaussian processes are widely used as surrogate models in Bayesian
optimization (BO)~\cite{jones1998efficient,snoek2012practical}, making
BO a natural setting for testing whether (rescaled) channel
QGPs are useful beyond regression. Here, rather than predicting an
observable over a fixed set of input states, the QGP posterior is used
to select new states for which the observable should be evaluated, allowing the model to guide the optimization of the observable
over a parametrized family of input states.
We note that such optimization is a common
task in quantum algorithms~\cite{cerezo2020variationalreview}. While usually the optimization goal is 
state-preparation under unitary dynamics,
optimization under open-system evolution is also of practical
interest~\cite{yoshioka2020variational}. 

BO is particularly useful when evaluations of the
objective function are expensive, and the optimization landscape is
non-convex~\cite{jones1998efficient}. Both features commonly arise when
optimizing expectation values over parametrized quantum
states~\cite{mcclean2018barren,anschuetz2022quantum}, and BO has already
been applied successfully to quantum observable optimization
~\cite{cheng2024quantum,jager2026provable}. The choice of the GP kernel can strongly
affect the efficiency of the optimization~\cite{snoek2012practical}.

More explicitly, we aim to  optimize
$\ell_{\CC}^{(\vec{t})}$ over the parameters $\vec{t}$. Starting from one (a few) initial evaluation(s), BO constructs a QGP surrogate for
the objective and leverages its posterior to sequentially propose the next parameter evaluation via an acquisition function. This function explicitly balances exploration (sampling regions of high model uncertainty) and exploitation (sampling optima of the mean prediction). Iteratively updating the surrogate with these new observations progressively refines the QGPs as much as needed to determine the optimum, thereby limiting the need for experimentally or computationally expensive channel evaluations.

We consider here a
two-parameter state-preparation problem for the non-equilibrium dynamics
of a spin chain, where Pauli observables provide natural probes of the
resulting many-body dynamics
~\cite{dutta2016anti,singh2021driven,kempa2026boundary}.
Specifically, we consider a chain of $n=10$ qubits and a family of initial pure spin-spiral product states
\begin{equation}
    \label{eq:spiral_states}
    \ket{\psi(\theta,q)}
    =
    \bigotimes_{j=0}^{n-1}
    \left(
        \cos\frac{\theta}{2}\ket{0}_j
        +
        e^{i q j}\sin\frac{\theta}{2}\ket{1}_j
    \right)\,,
\end{equation}
where $\vec{t} = (\theta,q)$, and the parameter ranges are 
\begin{equation}
    \theta \in [0,\pi],
    \qquad
    q \in [-\pi,\pi]\,.
\end{equation}
The parameter $\theta$ controls the local polarization of the initial state, while $q$ controls the pitch of the spiral phase profile. These states model exotic spin orders occurring in strongly-correlated materials~\cite{kumar2010spin}.     
We model the chain dynamics with Trotterized time evolution of an XXZ Hamiltonian, which is a paradigmatic model in the field of quantum many-body dynamics~\cite{singh2021driven,kempa2026boundary,cerezo2017factorization},
\begin{equation}
\label{eq:XXZ}
H_{XXZ}
    =
    J\sum_{j=0}^{n-2}
    \left(
        X_jX_{j+1}
        +
        Y_jY_{j+1}
        +
        \Delta Z_jZ_{j+1}
    \right)\,
\end{equation}
with a site- and time-dependent field term
\begin{equation}
 \label{eq:XXZ_1q}
    H_1
    =
    g_x\sum_{j=0}^{n-1}X_j
    +
    \sum_{j=0}^{n-1}h_j(\tau)Z_j\,,
\end{equation}
and additional single-qubit depolarizing noise. We note that to avoid a notation conflict we denote time  by $\tau$. The Hamiltonian parameters are $J=-1$, $\Delta=1$, and $g_x=0.8$, and we consider 20 Trotter steps with a time step $\delta \tau=0.05$. In addition, the longitudinal single-body terms  vary randomly in time. This is modeled by randomly and independently sampling each $h_j$ at each Trotter step once from an interval $[-0.9,0.9]$.  We model a coupling to an environment through  single-qubit depolarizing noise of strength $p=0.003$, which occurs after each two-qubit term in the Trotterized evolution operator.  This noisy dynamics defines the  channel~$\mathcal{C}$. 
Further implementation details
are provided in Appendix~\ref{app:spiral_benchmark}. 

\begin{figure*}[t]
\includegraphics[width=0.99\linewidth]{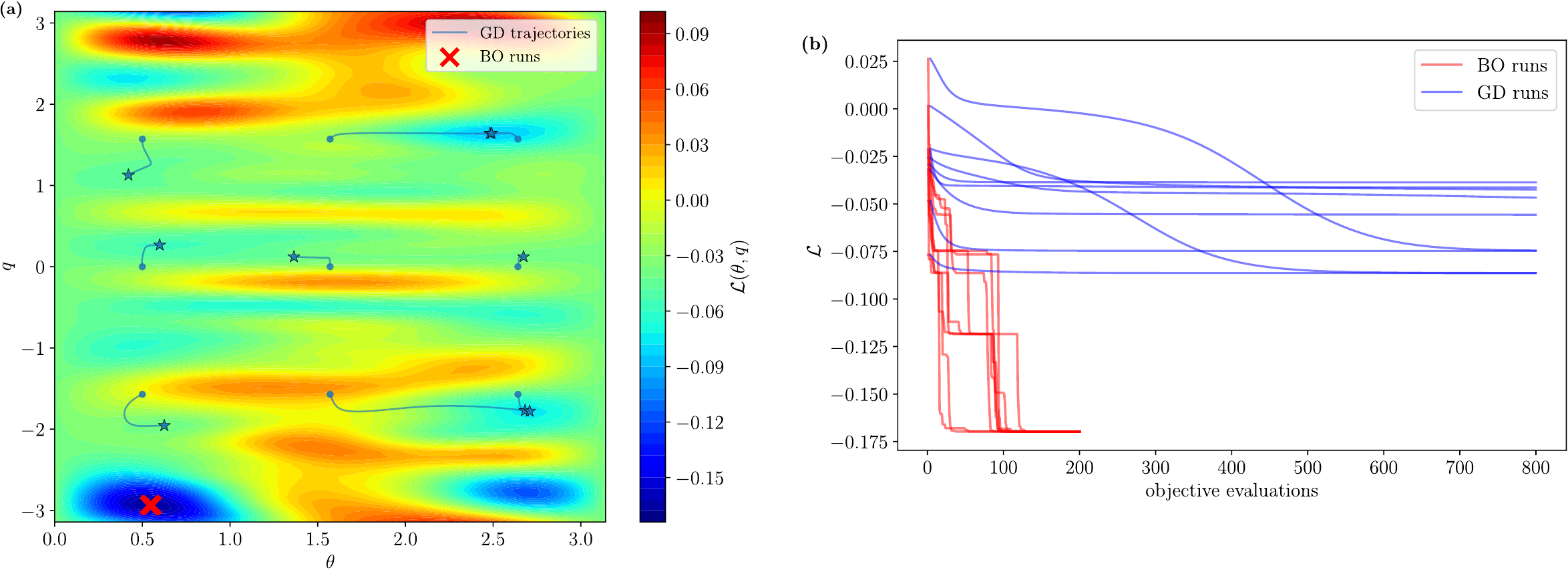}
\caption{
{\bf Bayesian optimization of noisy XXZ dynamics with channel GP surrogates.} Here we show BO of staggered magnetization (\ref{eq:M_stagg}) for the dynamics of a 10-qubit XXZ Hamiltonian with transverse and disordered longitudinal fields (\ref{eq:XXZ}, \ref{eq:XXZ_1q}) and single-qubit depolarizing noise. 
As a reference, we show gradient descent (GD) optimization. 
We optimize over a family of input spin-spiral states parametrized by parameters $\theta$ and $q$ (\ref{eq:spiral_states}). In (a),  the contour plot shows our loss function landscape (\ref{eq:spiral_loss}) on an 80-by-80 grid. The overlapping red  crosses show the best parameters found by 9 BO runs, indicating that all the runs converge to the same minimum. The BO uses channel QGPs with rescaled kernels as surrogate models.  The blue lines show trajectories  of 9  GD optimization runs. Each trajectory shows an initial point (circle) and an end point (star). The GD runs have the same initial points as the BO runs. Each GD (BO) run consists of $800$ ($100$) loss evaluations performed numerically using a full density matrix simulator.  In (b),  we plot the best loss value versus the number of loss evaluations for the BO (the red curves) and gradient descent (the blue lines) runs.
}
\label{fig:bo}
\end{figure*}

Our observable of interest is final-time staggered magnetization,
\begin{equation}
    \label{eq:M_stagg}
    M_{\rm stag}
    =
    \frac{1}{n}\sum_{j=0}^{n-1}(-1)^j Z_j\,.
\end{equation}
Its expectation value is calculated numerically in the infinite shot limit. 
Thus,  our loss function is
\begin{equation}
    \mathcal{L}(\theta,q)
    = \ell_{\CC}^{(\vec{t})} =
    \Tr\!\left[
        M_{\rm stag}\,
        \mathcal{C}\!\left(\dyad{\psi(\theta,q)}\right)
    \right]\,.
    \label{eq:spiral_loss}
\end{equation}
The spin-spiral states are product states, so their pairwise overlaps can
be evaluated efficiently. For the $10$-qubit system considered here, the
computational cost is therefore dominated by numerical evaluations of
$\mathcal{L}(\theta,q)$. The
resulting landscape, shown in Fig.~\ref{fig:bo} on an $80\times80$
uniform grid, is non-convex and contains extended regions with small
gradients. This provides a challenging optimization problem in which
reducing the number of loss evaluations is crucial for the optimization efficiency.

We perform Bayesian optimization runs from $9$ different initial points using the rescaled channel QGP
as the surrogate model. Each run consists of $100$ iterations, with one
loss evaluation performed at each iteration. The next parameters are
chosen by maximizing the expected improvement acquisition function \cite{jones1998efficient}. At
each iteration, the overall scale of the rescaled channel kernel is
determined from the data accumulated thus far. Further details
of the optimization procedure are provided in
Appendix~\ref{app:bo_impl}.

As shown in Fig.~\ref{fig:bo}, all $9$ BO runs locate the global minimum
identified from the loss landscape with high accuracy within $100$ loss
evaluations. We compare these results with gradient descent (GD) using finite-difference gradients and perform $9$ independent runs from the same initial points. All of the
GD runs fail to reach the global minimum despite using $800$ loss
evaluations. GD implementation details are provided
in Appendix~\ref{app:gd}.

\section{Summary and discussion}
\label{sec:discussion}

In this work, we generalized quantum Gaussian process regression from
unknown unitary evolutions to general quantum channels. The key step is
to place a prior over the channel itself. When no channel-specific
information is available, we take this prior to be the Lebesgue measure
over quantum channels and derive the corresponding QGP analytically. The
prior mean vanishes and the covariance is determined by the pairwise
overlaps of the input states, together with a dimension-dependent
scaling factor. This gives a closed-form channel QGP with a (hyper)parameter-free quantum kernel. When the channel acts on
$\mathcal{O}(\log n)$ qubits of a larger $n$-qubit system, the dimensional suppression remains only
inverse polynomial, allowing the channel QGP to be resolved with polynomial
shot resources even when the full quantum system is much larger.

Indeed, we performed numerical simulations aimed at demonstrating this regime. For a four-qubit
global channel, as well as for a $64$-qubit system with a local channel acting
only on a four-qubit subsystem, the channel QGP learns the
observable dependence from a small number of training states. In these
examples, the QGP reaches comparable prediction errors to row-PTM
learning with one to two orders of magnitude fewer shots. We also use
the same channel kernel for a channel implemented on IBM's quantum
computer \texttt{ibm\_boston}, where a data set of size $k=6$ is sufficient to reproduce the
measured dependence of the observable on the input state parameter. These examples
realize the regime identified by the scaling analysis, where the Stinespring-dilation dimension is small enough for efficient QGP learning in the presence of finite-shot noise.

When the Stinespring-dilation dimension grows, the Lebesgue prior becomes
increasingly conservative. A priori, if nothing is known about the
implemented channel, there is no reason
to expect non-zero channel observations. Indeed, averaging the observations over all channels
gives a vanishing mean, while the covariance in
Eq.~\eqref{eq:Lebesgue_kernel} is suppressed as the dimension grows. The
Lebesgue prior therefore assigns very little weight to learnable structure in the observable values in large systems. In an experiment, however, the
situation is different. In a controlled experiment or well-behaved quantum device, a reproducible
non-zero signal is a probe of implemented
dynamics, and the measured dependence on the input states provides
information about the particular channel being realized. Keeping the
dimensional scaling factor of this prior then amounts to assigning an exponentially small
prior scale to correlations that are directly supported by the data.
This motivates retaining the fidelity dependence derived from the Lebesgue
channel average while allowing the overall scale of the correlations to
be learned from the device.

The rescaled channel kernel is motivated by this distinction. We retain
the dependence on the input-state fidelity selected by the
Lebesgue theoretical calculation, while allowing its overall scale to be determined
from the data. The overlap remains physically meaningful since
it controls how strongly the Lebesgue prior correlates the observable values
associated with different input states. The fitted prefactor instead
relaxes the normalization imposed by the maximally uninformative channel
ensemble. In this sense, the rescaled QGP incorporates some trust in the
observed device response without discarding the state-dependent structure
obtained from the analytic calculation.

These two regimes are reflected in our numerical results. For the
four-qubit global channel, where the Lebesgue covariance can still be
resolved with a moderate shot count, fitting the overall scale has little effect
and the channel and rescaled kernels perform similarly. For the $64$-qubit
example with $n_A=64$, the Lebesgue dimensional prefactor exponentially
suppresses the covariance. As expected, increasing the shot budget over a practically feasible range fails to improve the channel QGP predictions. The
rescaled QGP, in contrast, improves systematically with the shot budget.
The fitted scale is therefore most useful when the measured device
response contains structure that is strongly suppressed by the
uninformed Lebesgue prior.

We also showcase the utility of channel QGPs in settings beyond regression. In the noisy XXZ state-preparation
problem, the channel QGP serves as a surrogate for Bayesian optimization, allowing us to find the optimal initial state more efficiently than in the case of standard optimization methods. We note that to provide a proof-of-principle feasibility demonstration, we perform the task numerically. This serves as a simplified numerical model of an experimental black box optimization, when the channel is genuinely unknown. At the same time, it serves as an example of  efficiency gains in the case when  the channel is known, but computing its  observations is expensive. Such a case occurs frequently when the channel aggregates the effects of many well-characterized channels, like for  a channel of a device-level quantum circuit implementation  built of well-calibrated and characterized quantum gates.

The remaining question is how to construct channel priors that encode
more of what is actually known about a physical application and how much we trust them. The Lebesgue
ensemble is appropriate when essentially no channel-specific information
is available, but experimentally relevant channels often come with
additional structure from locality, symmetries, calibrated noise, or
restricted system-environment couplings. The unitary setting already
shows that such structure can lead to provable and scalable QGPs, as occurs for matchgate evolutions
~\cite{jager2026provable}. For quantum channels, the corresponding
problem is to determine which experimentally justified assumptions are
sufficient to derive informative priors whose kernels remain efficiently resolvable, as necessary for channel QGPs that are both provable and scalable.

The rescaled channel kernel provides a heuristic step in this direction. It does
not correspond to a Lebesgue channel ensemble derived here, but it
demonstrates how the QGP framework can be used beyond the priors that are
derived analytically while retaining the physical
information contained in the analytic calculation. A natural next step is
therefore to derive device-informed channel ensembles that preserve this
useful correlation structure while replacing the uniform Lebesgue
normalization by a scale consistent with experimentally available
knowledge.

\section*{Artificial Intelligence Disclosure}
The authors acknowledge the use of Claude Opus 5, ChatGPT 5.6 Sol, and Gemini 3.1 Pro for code development and for writing and reviewing the manuscript. All generated results were reviewed and validated by the authors.

\section*{Acknowledgments}
JJ acknowledges support from the Natural Sciences and Engineering Research Council (NSERC) of Canada, specifically the NSERC CREATE in Quantum Computing Program (grant number 543245).  PC and YK acknowledge support  by the National Science Centre (NCN), Poland under project 2022/47/D/ST2/03393. AM acknowledges support from the Priority Research Area Digiworld under the program Excellence Initiative – Research University at the Jagiellonian University in Kraków. Part of this work was carried out while affiliated with the Institute of Theoretical Physics and the Mark Kac Center for Complex Systems Research, Jagiellonian University, Kraków, Poland, and part while affiliated with the $\Phi$-lab, European Space Agency (ESA/ESRIN), Frascati, Italy. DGM acknowledges financial support from the European Research Council (ERC) via the Starting grant q-shadows (101117138) and from the Austrian Science Fund (FWF) via the SFB BeyondC (10.55776/FG7). MC acknowledges support from Los Alamos National Laboratory (LANL) ASC Beyond Moore’s Law project. This work was also supported by the Quantum Science Center (QSC), a National Quantum Information Science Research Center of the U.S. Department of Energy (DOE). This research used quantum computing resources provided by the LANL Institutional Computing Program, which is supported by the U.S. DOE National Nuclear Security Administration under Contract No. 89233218CNA000001. We acknowledge the use of IBM Quantum services for this work. The views expressed are those of the authors, and do not reflect the official policy or position of IBM or the IBM Quantum team.
\\ ESA classification: UNCLASSIFIED -- Releasable to the Public.

\bibliography{quantum,new_references}

\appendix

\section{Stability for finite-shot kernel estimates}
\label{app:finite_shot_stability}

The analysis in Sec.~\ref{sec:channel_qgp:scalable} concerns the finite-shot uncertainty in the training
values, which enters the posterior through $\Gamma$. However, one must also note that the covariance
matrix is estimated from
finite-shot overlap measurements, and this uncertainty must be taken into account. As such, we consider how errors in the
estimated covariance matrix propagate through the inverse appearing in
the QGP posterior.

Let
\begin{equation}\label{eq:stability_A_def}
    A=\Sigma+\Gamma,
\end{equation}
and consider an estimated covariance matrix with additive error $\Delta$,
such that
\begin{equation}
    \widehat{A}=A+\Delta.
\end{equation}
We assume throughout that the estimated covariance matrix has been symmetrized so that $\Delta$ is symmetric and $\widehat{A}$ remains Hermitian.
We denote by $\lambda_{\min}(X)$ the smallest eigenvalue of $X$ and by
$\|X\|_{\rm op}$ its operator norm. To isolate the effect of the
covariance-matrix error, we keep $\vec{m}$ and $\vec{y}$ fixed. If
\begin{equation}\label{eq:stability_A_norm_eigval_ineq}
    \|\Delta\|_{\rm op}<\lambda_{\min}(A),
\end{equation}
then
\begin{equation}
    \label{eq:inverse_stability}
    \|\widehat{A}^{-1}-A^{-1}\|_{\rm op}
    \leq
    \frac{\|\Delta\|_{\rm op}}
    {\lambda_{\min}(A)
    \left(
        \lambda_{\min}(A)-\|\Delta\|_{\rm op}
    \right)}.
\end{equation}
Consequently, the error in the posterior mean satisfies
\begin{equation}
    \left|
        \widehat{\mu}_{P}-\mu_{P}
    \right|
    \leq
    \|\vec{m}\|_2\|\vec{y}\|_2
    \frac{\|\Delta\|_{\rm op}}
    {\lambda_{\min}(A)
    \left(
        \lambda_{\min}(A)-\|\Delta\|_{\rm op}
    \right)}.
    \label{eq:posterior_stability}
\end{equation}
In the regime
$\|\Delta\|_{\rm op}\ll\lambda_{\min}(A)$, this reduces to
\begin{equation}
    \left|
        \widehat{\mu}_{P}-\mu_{P}
    \right|
    \in
    \mathcal{O}\left(
        \|\vec{m}\|_2\|\vec{y}\|_2
        \frac{\|\Delta\|_{\rm op}}
        {\lambda_{\min}(A)^2}
    \right).
\end{equation}
The proof of this result is provided in the next section. We also refer the reader to Ref.~\cite{xu2026active}, where related
inverse-stability arguments have been used in the analysis of finite-shot quantum kernel estimation in Gaussian process regression.

The previous bound shows that finite-shot errors in the covariance matrix remain
controlled when $\|\Delta\|_{\rm op}$ is small compared with
$\lambda_{\min}(\Sigma+\Gamma)$. Conversely, errors in the estimated
kernel can be strongly amplified when the regularized covariance matrix
has small eigenvalues. This condition is independent of the particular
channel ensemble and applies to both the analytic and rescaled kernels.
We note that the finite-shot effects considered in this appendix are distinct from the scaling phenomenon discussed in Secs.~\ref{sec:channel_qgp:scalable} and \ref{sec:rescaled_kernel}.
The present analysis concerns errors in estimating the kernel entries, while the learnability limitations of the channel kernel arise from the relative magnitude of the observation noise and the exact kernel scale, that is $d$ and $1/\eta$, for the channel kernel and the rescaled channel  kernel respectively.

\subsection{Finite-shot stability proofs}
\label{app:finite_shot}

Given Eqs.~\eqref{eq:stability_A_def}--\eqref{eq:stability_A_norm_eigval_ineq},
we use the resolvent identity,
\begin{equation}
    \widehat{A}^{-1}-A^{-1}
    =
    -\widehat{A}^{-1}\Delta A^{-1},
\end{equation}
which implies
\begin{equation}
    \|\widehat{A}^{-1}-A^{-1}\|_{\rm op}
    \leq
    \|\widehat{A}^{-1}\|_{\rm op}
    \|\Delta\|_{\rm op}
    \|A^{-1}\|_{\rm op}.
\end{equation}
Since $A$ is positive definite,
\begin{equation}
    \|A^{-1}\|_{\rm op}
    =
    \frac{1}{\lambda_{\min}(A)}.
\end{equation}
Furthermore, Weyl's inequality gives
\begin{equation}
    \lambda_{\min}(\widehat{A})
    \geq
    \lambda_{\min}(A)-\|\Delta\|_{\rm op},
\end{equation}
and therefore
\begin{equation}
    \|\widehat{A}^{-1}\|_{\rm op}
    \leq
    \frac{1}{
        \lambda_{\min}(A)-\|\Delta\|_{\rm op}
    }.
\end{equation}
Combining the two bounds yields
\begin{equation}
    \|\widehat{A}^{-1}-A^{-1}\|_{\rm op}
    \leq
    \frac{
        \|\Delta\|_{\rm op}
    }{
        \lambda_{\min}(A)
        \left(
            \lambda_{\min}(A)-\|\Delta\|_{\rm op}
        \right)
    }.
\end{equation}

For the posterior mean
\begin{equation}
    \mu_P
    =
    \vec{m}^{\,T}A^{-1}\vec{y},
\end{equation}
the induced perturbation is
\begin{equation}
    \widehat{\mu}_P-\mu_P
    =
    \vec{m}^{\,T}
    \left(
        \widehat{A}^{-1}-A^{-1}
    \right)
    \vec{y},
\end{equation}
from which
\begin{equation}
    |\widehat{\mu}_P-\mu_P|
    \leq
    \|\vec{m}\|_2
    \|\vec{y}\|_2
    \|\widehat{A}^{-1}-A^{-1}\|_{\rm op}
\end{equation}
follows. Thus,
\begin{equation}
    |\widehat{\mu}_P-\mu_P|
    \leq
    \|\vec{m}\|_2
    \|\vec{y}\|_2
    \frac{
        \|\Delta\|_{\rm op}
    }{
        \lambda_{\min}(A)
        \left(
            \lambda_{\min}(A)-\|\Delta\|_{\rm op}
        \right)
    }.
\end{equation}
For
$\|\Delta\|_{\rm op}\ll\lambda_{\min}(A)$, this gives
\begin{equation}
    |\widehat{\mu}_P-\mu_P|
    \in
    \mathcal{O}\left(
        \|\vec{m}\|_2
        \|\vec{y}\|_2
        \frac{
            \|\Delta\|_{\rm op}
        }{
            \lambda_{\min}(A)^2
        }
    \right).
\end{equation}

\section{Physical systems and simulation details}

\subsection{Ising dynamics of a system with an environment}
\label{app:Ising_impl}

As a test case, in
Sections~\ref{sec:channel_GPR_experiments}
and~\ref{sec:heur_GPR_numerics}, we consider a quantum system
$\HC_S=(\mathbb{C}^{2})^{\otimes n}$ and an environment
$\HC_E=(\mathbb{C}^{2})^{\otimes n_{\rm env}}$, with
$(n,n_{\rm env})\in\{(4,4),(64,4),(64,64)\}$.

The system is prepared  in a state 
\begin{equation}
\quad \ket{\psi_i(t)} =\bigotimes_{j=0}^{n-1} e^{-i \phi(j,t) Y_j} e^{-i \phi(j,t) X_j}\ket{0}_j,
\end{equation}
with
\begin{equation}
\phi(j,t) = \frac{\pi}{8} f(j) g(t), 
\end{equation}
and
\begin{equation}
f(j) = e^{-a(j-j_0)^2}, \quad g(t) = \sin(2\pi t)+\cos(4\pi t)/2.
\label{eq:rho0_QIs_app_envelope} 
\end{equation}
For $(n,n_{\rm env})=(4,4)$ we choose $j_0=2$, $a=1$, for $(n,n_{\rm env})=(64,4)$ we set $j_0=60$, $a=1/4$, and for $(n,n_{\rm env})=(64,64)$ we have $j_0=32$, $a=1/256$.

After the initial state preparation, the system is acted on by a unitary operator $U_S$ implemented as an instance of Trotter decomposition of transverse-field Ising dynamics 
\begin{equation}
\Big[
    e^{-i\delta \tau H_{Y}}
     e^{-i\delta \tau H_{X}}
    e^{-i\delta \tau\, \big(H_{ZZ}+H_{Z}\big)}
  \Big]^{N_{\rm step}}, 
  \label{eq:Ising_dynamics}
\end{equation}
with 
\begin{equation}
H_{ZZ} = J\sum_{j=n_i}^{n_{f}-2}Z_j Z_{j+1},
\label{eq:HZZ}
\end{equation}
and
\begin{equation}
H_{X} = h \sum_{j=n_i}^{n_f-1}   X_j, \, H_{Y} = h \sum_{j=n_i}^{n_f-1}   Y_j, \, H_{Z} = h \sum_{j=n_i}^{n_f-1}   Z_j.
\label{eq:H1q}
\end{equation}
Here, we assume that system indices are labeled by $i\in\{0,\dots,n-1\}$ and the environment indices are numbered by 
$i\in\{n,\dots,n+n_{\rm env}-1\}$. 
For $U_S$, we set $n_i=0$, $n_f=n$, $J=1$, $h=1/3$, and
$N_{\rm step}=n/2$. Furthermore, for the numerical experiments with $(n,n_{\rm env})=(4,4)$ and $(n,n_{\rm env})=(64,4)$ we choose $\delta \tau=1/N_{\rm step}$, and   $\delta \tau=0.2/N_{\rm step}$, respectively. For $(n,n_{\rm env})=(64,64)$, we have $\delta \tau=1/N_{\rm step}$, and for the real-hardware implementation we choose $\delta \tau=0.02/N_{\rm step}$.

Next, the system is coupled to the environment prepared in a state $\ket{0}^{\otimes n_{\rm env}}$ by a 
unitary $U_{SE} \in \BC( \HC_S\otimes \HC_E)$.
The unitary $U_{SE}$ is implemented as the  Trotterized time evolution (\ref{eq:Ising_dynamics}). 
We set $J=1$  and $n_f=n+n_{\rm env}$. 
For $(n,n_{\rm env})=(4,4)$, and $(n,n_{\rm env})=(64,64)$ we have  $n_i=0$, while for $(n,n_{\rm env})=(64,4)$ we choose $n_i=60$. 
For  $(n,n_{\rm env})=(4,4)$ and $(n,n_{\rm env})=(64,4)$ we have
$N_{\rm step}=4$, while for $(n,n_{\rm env})=(64,64)$ we use $N_{\rm step}=64$.  In the case of numerical experiments with  $(n,n_{\rm env})=(4,4)$ and  $(n,n_{\rm env})=(64,64)$ we set $\delta \tau=0.5/N_{\rm step}$ and $h=1/3$,  while for the hardware implementation we choose $\delta \tau=0.08/N_{\rm step}$ and $h=0.1$. For  $(n,n_{\rm env})=(64,4)$, we use $h=1/3$ and  $\delta \tau=0.75/N_{\rm step}$.  
The observable of interest  for  $(n,n_{\rm env})=(4,4)$  is $\langle X_1 \rangle$, for $(n,n_{\rm env})=(64,4)$  it is  $\langle X_{60} \rangle$, and for $(n,n_{\rm env})=(64,64)$  we learn $\langle X_{31} \rangle$.

In the numerical experiments we choose the input  states
as 
\begin{equation}
\rho(t) =  U_S\dyad{\psi_i(t)} U_S^{\dag}
\end{equation}
and the channel as
\begin{equation}
  \CC(\rho) \;=\; \Tr_E\!\left[\,U_{SE}\big(\rho\otimes \dyad{\vec{0}}\big)U_{SE}^\dagger\,\right]\,.
\end{equation}
In the hardware implementation, the input  states
are 
\begin{equation}
\rho(t) =  \dyad{\psi_i(t)}
\end{equation}
and the channel is
\begin{equation}
  \CC(\rho) \;=\; \Tr_E\!\left[\,U_{SE}U_S\big(\rho\otimes \dyad{\vec{0}}\big)U_S^{\dag}U_{SE}^\dagger\,\right]\,.
\end{equation}
We note that this division results in different covariances for the subsystem kernel (\ref{eq:Lebesgue_kernel_subsystem}).
We note that our observable and channel choices imply that  effectively the channel has $n_B=1$. For  $(n,n_{\rm env})=(64,64)$,  we have $n_A=64$, while $n_A=4$ otherwise.

\paragraph*{Numerical tensor network simulation.}
When coupling the $64$-qubit system to a $64$-qubit environment, the channel scales extensively with the system size. Due to the choice of Ising dynamics, both the observable of interest and the input-state overlaps can be computed classically using matrix product state
methods~\cite{schollwock2011density}. We use these simulations to generate the data required for QGP regression and to evaluate the prediction accuracy.

\paragraph*{IBM hardware implementation.}
 For the purpose of the IBM implementation, we decompose the state preparation and the  time evolutions to single-qubit $R_X$, $R_Z$, $R_Y$ and two-qubit $R_{ZZ}$ gates. The two-qubit gates are subsequently decomposed to $R_Z$ and $CNOT$ gates. The resulting circuit is run on the IBM device \texttt{ibm\_boston}. To utilize the large qubit count of the device,  we run a circuit obtained by dividing the device graph to 19 disjoint subgraphs and executing the 8-qubit time evolutions on each of the subgraphs simultaneously. For each $t$ we gather 1600 shots. This choice implies that effectively we gather $19\cdot 1600=34400$ shots per each $t$ value.

\subsection{Spiral state preparation under noisy XXZ dynamics for Bayesian optimization demonstration}
\label{app:spiral_benchmark}

Here, we describe in detail the state preparation for Bayesian optimization demonstration from Section~\ref{sec:bo_vs_gd}. 
The problem setting is the following. We consider a chain of $n$ qubits with Hilbert space
$
    \HC = (\mathbb{C}^2)^{\otimes n}%
$.
In the numerical experiment shown in Fig.~\ref{fig:bo} we set $n=10$. We denote by $X_j$, $Y_j$, and $Z_j$ the Pauli operators acting on site $j$. 
We  define a parametrized spin-spiral product state
\begin{equation}
    \ket{\psi(\theta,q)}
    =
    \bigotimes_{j=0}^{n-1}
    \left(
        \cos\frac{\theta}{2}\ket{0}_j
        +
        e^{i q j}\sin\frac{\theta}{2}\ket{1}_j
    \right)\,,
    \label{eq:spiral_state_appendix}
\end{equation}
and denote the associated density matrix as
\begin{equation}
    \rho(\theta,q)=\dyad{\psi(\theta,q)}\,.
\end{equation}
The parameter $\theta$ controls the relative amplitude between the basis states $\ket{0}$ and $\ket{1}$, while $q$ controls the spatial winding of the phase, namely how fast the latter changes over the chain. This family of quantum states is simple to prepare, but after undergoing the noisy interacting dynamics described below it gives rise to a nontrivial two-dimensional magnetization landscape.

We implement noisy dynamics as an XXZ Hamiltonian with transverse and longitudinal fields and depolarizing noise. At Trotter step $m$ we set
\begin{equation}
    H^{(m)}
    =
    H_{\rm even}
    +
    H_{\rm odd}
    +
    H_1^{(m)}\,,
\end{equation}
where the two-body terms are split into even and odd bonds,
\begin{align}
    H_{\rm even}
    &=
    J\sum_{\substack{j=0\\ j\,{\rm even}}}^{n-2}
    \left(
        X_jX_{j+1}
        +
        Y_jY_{j+1}
        +
        \Delta Z_jZ_{j+1}
    \right)\,,
    \\
    H_{\rm odd}
    &=
    J\sum_{\substack{j=0\\ j\,{\rm odd}}}^{n-2}
    \left(
        X_jX_{j+1}
        +
        Y_jY_{j+1}
        +
        \Delta Z_jZ_{j+1}
    \right)\,,
\end{align}
while the one-body term is
\begin{equation}
    H_1^{(m)}
    =
    g_x\sum_{j=0}^{n-1}X_j
    +
    \sum_{j=0}^{n-1}h_j^{(m)}Z_j\,.
\end{equation}
We set
\begin{equation}
    J=-1\,,
    \qquad
    \Delta=1\,,
    \qquad
    g_x=0.8\,.
\end{equation}
The longitudinal fields are chosen by  sampling independently from a uniform distribution for each Trotter step
\begin{equation}
    h_j^{(m)} \sim {\rm Unif}[-s,s]\,,
    \qquad
    s=0.9\,.
\end{equation}
We sample a single disorder realization $\{h_j^{(m)}\}$, which is used for all optimization runs.

We use
\begin{equation}
    N_{\rm step}=20,
    \qquad
    \Delta \tau=0.05,
\end{equation}
corresponding to a total evolution time
\begin{equation}
    T=N_{\rm step}\delta \tau=1.
\end{equation}
Each time step is implemented using a symmetric second-order product formula~\cite{suzuki1976generalized, lloyd1996universal}. 
Particularly, defining
\begin{align}
    U_1^{(m)}(\tau)
    &=
    e^{-i\tau H_1^{(m)}}\,,
    \\
    U_{\rm even}(\tau)
    &=
    e^{-i\tau H_{\rm even}}\,,
    \\
    U_{\rm odd}(\tau)
    &=
    e^{-i\tau H_{\rm odd}}\,,
\end{align}
the noiseless unitary for step $m$ is
\begin{equation}
    \label{eq:spiral_second_order_trotter}
    \begin{split}
    &U_{\delta \tau}^{(m)}
    = \\
    &\;\; U_1^{(m)}(\tfrac{\delta \tau}{2})\,
    U_{\rm even}(\tfrac{\delta \tau}{2})\,
    U_{\rm odd}(\delta \tau)\,
    U_{\rm even}(\tfrac{\delta \tau}{2})\,
    U_1^{(m)}(\tfrac{\delta \tau}{2})\,. 
    \end{split}
\end{equation}
Notice that the even and odd two-body layers are products of commuting nearest-neighbor gates and can hence be implemented as brickwork layers of gates.

To model imperfect hardware, we include homogeneous single-qubit depolarizing noise with strength $p=0.003$. For a single qubit,
\begin{equation}
    \mathcal{D}_{p}(\rho)
    =
    \left(1-\frac{3p}{4}\right)\rho
    +
    \frac{p}{4}
    \left(
        X\rho X + Y\rho Y + Z\rho Z
    \right)\,.
    \label{eq:depolarizing_appendix}
\end{equation}
We denote the corresponding $n$-qubit product channel by
\begin{equation}
    \mathcal{D}_{p}^{(n)}
    =
    \mathcal{D}_{p}^{\otimes n}\,.
\end{equation} 
Denoting by
\begin{equation}
    \mathcal{U}_A(\rho)=A\rho A^\dagger
\end{equation}
the unitary channel associated with $A$, the noisy channel for one Trotter step is then
\begin{equation}
\begin{split}
    \Phi^{(m)}
    =
    &\,
    \mathcal{U}_{U_1^{(m)}(\delta \tau/2)}
    \circ
    \mathcal{D}_{p}^{(n)}
    \circ
    \mathcal{U}_{U_{\rm even}(\delta \tau/2)}
    \circ
    \mathcal{D}_{p}^{(n)}
    \circ
    \mathcal{U}_{U_{\rm odd}(\delta \tau)}
    \\
    &\,
    \circ
    \mathcal{D}_{p}^{(n)}
    \circ
    \mathcal{U}_{U_{\rm even}(\delta \tau/2)}
    \circ
    \mathcal{U}_{U_1^{(m)}(\delta \tau/2)}\,.
\end{split}
\label{eq:spiral_noisy_step}
\end{equation}
The full noisy channel is hence
\begin{equation}
    \Phi_T
    =
    \Phi^{(N_{\rm step})}
    \circ \cdots \circ
    \Phi^{(2)}
    \circ
    \Phi^{(1)}\,.
    \label{eq:spiral_full_channel}
\end{equation}

The observable that defines the loss landscape is the staggered magnetization
\begin{equation}
    M_{\rm stag}
    =
    \frac{1}{n}\sum_{j=0}^{n-1}(-1)^j Z_j\,.
\end{equation}
The loss function is then
\begin{equation}
    \mathcal{L}(\theta,q)
    =
    \Tr\!\left[
        M_{\rm stag}\,
        \Phi_T\!\left(\rho(\theta,q)\right)
    \right]\,.
    \label{eq:spiral_loss_appendix}
\end{equation}
The optimization problem is thus given by
\begin{equation}
    \min_{(\theta,q)\in\Omega}
    \mathcal{L}(\theta,q)\,.
\end{equation}
We evaluate the loss function numerically using a full density matrix simulator. 
The Bayesian optimization protocol is specified in Appendix~\ref{app:bo_impl}.

\section{Algorithmic and application implementation details}

\subsection{Channel quantum Gaussian process implementation}
\label{app:GPR_impl}

The QGP regression is implemented through the Python package \texttt{sklearn} \cite{pedregosa2011scikit}, using the channel and the rescaled channel kernels. For the channel kernel, as defined non-asymptotically in Eq.~\eqref{eq:Lebesgue_kernel_finite_n}, no kernel parameters are fitted. 
For the rescaled channel kernel, the overall prefactor $\eta$, as defined in Eq.~\eqref{eq:rescaled_channel_kernel}, is determined from the training data by
maximizing the GP marginal likelihood over $\eta \in [1/d,1]$ via L-BFGS-B \cite{byrd1995limited} (with 10 restarts). In the Bayesian
optimization experiments, $\eta$ is re-optimized after each new loss
evaluation using the data accumulated up to that iteration. 

For finite-shot training kernel matrix estimates\footnote{We assume the estimates are symmetrized, achieved either by symmetric sampling or by replacing the estimate $\hat{\Sigma}$ with $(\hat{\Sigma} + \hat{\Sigma}^\top)/2$.} that are not positive semidefinite (PSD), a correction technique based on the Wigner semicircle law \cite{jager2026provable} is applied. This shifts the kernel diagonal by the semicircle bound, or, if insufficient to restore a PSD matrix, by the maximum negative eigenvalue magnitude
\begin{equation}
    \lambda_{\rm PSD} = \max\left\lbrace 2 \sqrt{k} \sigma_{\kappa}, \; \lvert \lambda_{\rm min} \rvert \right\rbrace,
\end{equation}
where $\sigma^2_{\kappa}$ and $\lambda_{\rm min}$ denote the sampling variance of a kernel entry and the lowest negative training kernel eigenvalue, respectively. 
Note that $\lambda_{\rm PSD}$ is directly interpretable as an increase in observation noise, i.e., we replace $\Gamma$ by $\Gamma + \lambda_{\rm PSD} \id$.

\subsection{Bayesian optimization implementation}
\label{app:bo_impl}

We now specify the Bayesian optimization protocol applied to the system and loss function in Appendix~\ref{app:spiral_benchmark}. The 9 Bayesian optimization runs are initialized using the $3\times3$ grid
\begin{equation}
    \mathcal{I}_0
    =
    \{0.5,\pi/2,\pi-0.5\}
    \times
    \{-\pi/2,0,\pi/2\}\,.
    \label{eq:bo_gd_initial_grid_appendix}
\end{equation}
For Bayesian optimization, we use a Gaussian process surrogate with the
rescaled channel kernel.
The optimization task is carried out over the domain
\begin{equation}
    \Omega
    =
    [0,\pi]\times[-\pi,\pi]\,,
\end{equation}
with coordinates
\begin{equation}
    \vec{t}=(\theta,q)\,.
\end{equation}
At iteration $r$, the $r$ loss evaluations
performed so far form the data set
\begin{equation}
    \mathcal{D}_r
    =
    \{(\theta_a,q_a,\mathcal{L}(\theta_a,q_a))\}_{a=1}^{r}.
\end{equation}
The data are used to compute the surrogate posterior mean
$\mu_r(\theta,q)$ and standard deviation $\sigma_r(\theta,q)$. The
overall kernel scale $\eta$ is reoptimized using $\mathcal{D}_r$ at
each iteration. The parameters of the next loss evaluation are chosen
by maximizing the expected-improvement acquisition
function~\cite{jones1998efficient,snoek2012practical}. For loss minimization,
the expected improvement is
\begin{equation}
\begin{split}
    {\rm EI}_r(\vec{t})
    &=
    \left(
        \mathcal{L}_{\min}^{(r)}-\mu_r(\vec{t})
    \right)
    \Phi_{\rm N}\!\left(z_r(\vec{t})\right) \\
    &\quad+
    \sigma_r(\vec{t})\,
    \phi_{\rm N}\!\left(z_r(\vec{t})\right)\,,
    \end{split}
    \label{eq:expected_improvement_appendix}
\end{equation}
given the best loss value found thus far (iteration $r$)
\begin{equation}
    \mathcal{L}_{\min}^{(r)}
    =
    \min_{a \in \lbrace 1, \ldots, r \rbrace }\mathcal{L}(\theta_a, q_a)\,,
\end{equation}
and
\begin{equation}
    z_r(\vec{t})
    =
    \frac{\mathcal{L}_{\min}^{(r)}-\mu_r(\vec{t})}{\sigma_r(\vec{t})}\,, \quad \vec{t} = (\theta, q),
\end{equation}
and $\Phi_{\rm N}$ and $\phi_{\rm N}$ are the cumulative distribution function and probability density, respectively, of a standard normal random variable. If $\sigma_r(\vec{t})=0$, we set ${\rm EI}_r(\vec{t})=0$. The next point to query is then
\begin{equation}
    \vec{t}_{r+1}
    =
    \arg\max_{\vec{t}\in\Omega}
    {\rm EI}_r(\vec{t})\,.
\end{equation}
Thus the total number of loss evaluations for a BO run with $N_{\rm BO}$ total queries is
\begin{equation}
    C_{\rm BO}
    =
    N_{\rm BO}.
\end{equation}

\section{Baseline methods}

\subsection{Row Pauli transfer matrix learning baseline}
\label{app:PTM}

A straightforward alternative approach to our channel QGP regression is  learning of  a row of a Pauli transfer matrix (PTM) for a Pauli observable of interest $O$.  We assume here $\CC:\BC((\C^2)^{\otimes n_A})\rightarrow\BC((\C^2)^{\otimes n_B})$.
The PTM elements corresponding to the observable of interest are 
\begin{equation}
R_{j} = \Tr[\CC(P_j)O]/2^{n_A},
\end{equation}
 where $j$ numbers $4^{n_A}$ Paulis that span  $\BC((\C^2)^{\otimes n_A})$. This basis can be constructed from tensor products of single-qubit Paulis $I$, $X$, $Y$, $Z$.  
 The  elements of $R$  can be estimated with two copies of the system qubit register $a$, $b$  and an initial state of Bell pairs of the qubits across the registers
\begin{align}
\ket{\Phi} = \frac{1}{2^{n_A/2}} \sum_{i_0=0}^{1} \dots \sum_{i_{n_A-1}=0}^{1} |i_0\rangle_{0,a}|i_0\rangle_{0,b} \dots \\
\dots|i_{n_A-1}\rangle_{n_A-1,a}|i_{n_A-1}\rangle_{n_A-1,b}. \nonumber
\end{align}
The channel $\mathcal{C}$ is applied to the second register of $\ket{\Phi}$ and finally $P_j^\top\otimes O$ is measured, as 
\begin{equation}
\Tr[\CC(P_j)O]/2^{n_A} \Tr \big[ (  P_j^\top \otimes O) (\mathbb{I} \otimes \mathcal{C}  ) \ket{\Phi} \big]. 
\end{equation}

To predict $ \Tr[\CC(\rho(t))O]$ we need to perform full state tomography of $\rho(t)$. This requires measurements of all $P_j\neq\mathbb{I}\in  \BC((\C^2)^{\otimes n_A})$ for the state, which determine its decomposition in the Pauli basis,
\begin{equation}
\rho(t) = \sum_j \frac{b_j}{2^{n_A}} P_j, \quad b_j  =  \Tr[\rho(t) P_j].
\end{equation}
Having this, we estimate
\begin{equation}
\Tr[\CC(\rho(t))O] = \sum_j R_{j}b_j.
\end{equation}
We see that in the row-PTM approach we need to perform $2\cdot4^{n_A}-1$ measurements for a single prediction. 

In Section~\ref{sec:channel_GPR_numerics}, we perform row-PTM learning  for a subsystem and a system  of $n_A=4$  qubits with the channel of interest implemented by a unitary coupling it to a $4$-qubit environment.
First, we investigate how a  division of shots between the row learning  and the state tomography  affects the method performance, using the 4-qubit system as a test case. To investigate that, we estimate the row $\{R_j\}$ using $N_s^{(1)}=i C$ shots and $\{b_j\}$  using $N_s^{(2)}=(20-i) C$ shots,  where $i\in \{1,2,\dots,19\}$, and $C\in\{2.56\cdot10^4,2.56\cdot10^6,2.56\cdot10^8\}$. We divide the shots equally between the row elements. For  the  $b$ coefficient estimation, we assign $\lfloor{N_s^{(2)}/(4^{n_A}-1)\rfloor}$ to each $b_j$, and divide randomly the remaining  shots among the coefficients. Hence, including the cost of training, we need in total $N_s^{(p)}=N_s^{(1)} + N_s^{(2)}$ shots for a single  prediction.

\begin{figure}[t!]
\includegraphics[width=.999\linewidth]{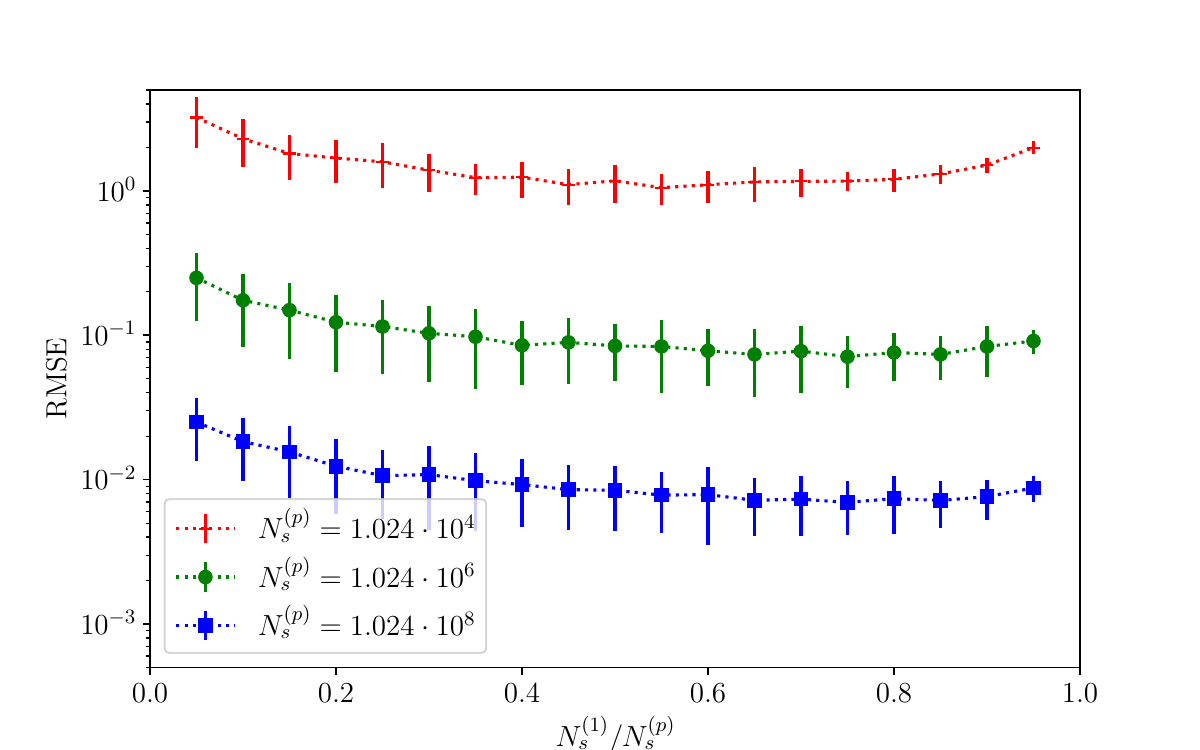}
\caption{{\bf Performance of the row-PTM  learning  versus shot distribution ratio used for the row-PTM learning. }  Here, we use a setup from Fig.~\ref{fig:GPR_Haar} with $n=n_A=4$.  We plot root mean squared error (RMSE) versus a ratio of the row learning shot cost  $N_s^{(1)}$  to the  total shots required for a prediction $N_s^{(p)}$. Here $N_s^{(p)}$ includes both shots for  the row learning and state tomography of the initial state. To account for randomness due to the shot noise, for each shot division we compute RMSE $100$ times, with each repetition using a different  instance of numerically simulated shot noise. In the  plot, for each ratio, we show the  mean (the marker) and the standard deviation (the error bar) of these RMSE estimates.}
\label{fig:PTM_hyper}
\end{figure}

For each pair of $N_s^{(1)}$ and $N_s^{(2)}$ values, we quantify the method performance by RMSE averaged over 100 shot noise instances. We gather the results in  Fig.~\ref{fig:PTM_hyper}. We observe that the error depends weakly on the shot division and is minimal for  $N_s^{(1)}/N_s^{(p)}\approx2/3$. Taking that into account, in the main text implementation from Fig.~\ref{fig:GPR_Haar}, we use $N_s^{(1)}/N_s^{(p)}\approx0.67$, with $ 2.56\cdot10^3 \le N_s^{(p)} \le 7.84 \cdot10^7$.

\begin{figure*}[htb!]
\includegraphics[width=.99\linewidth]{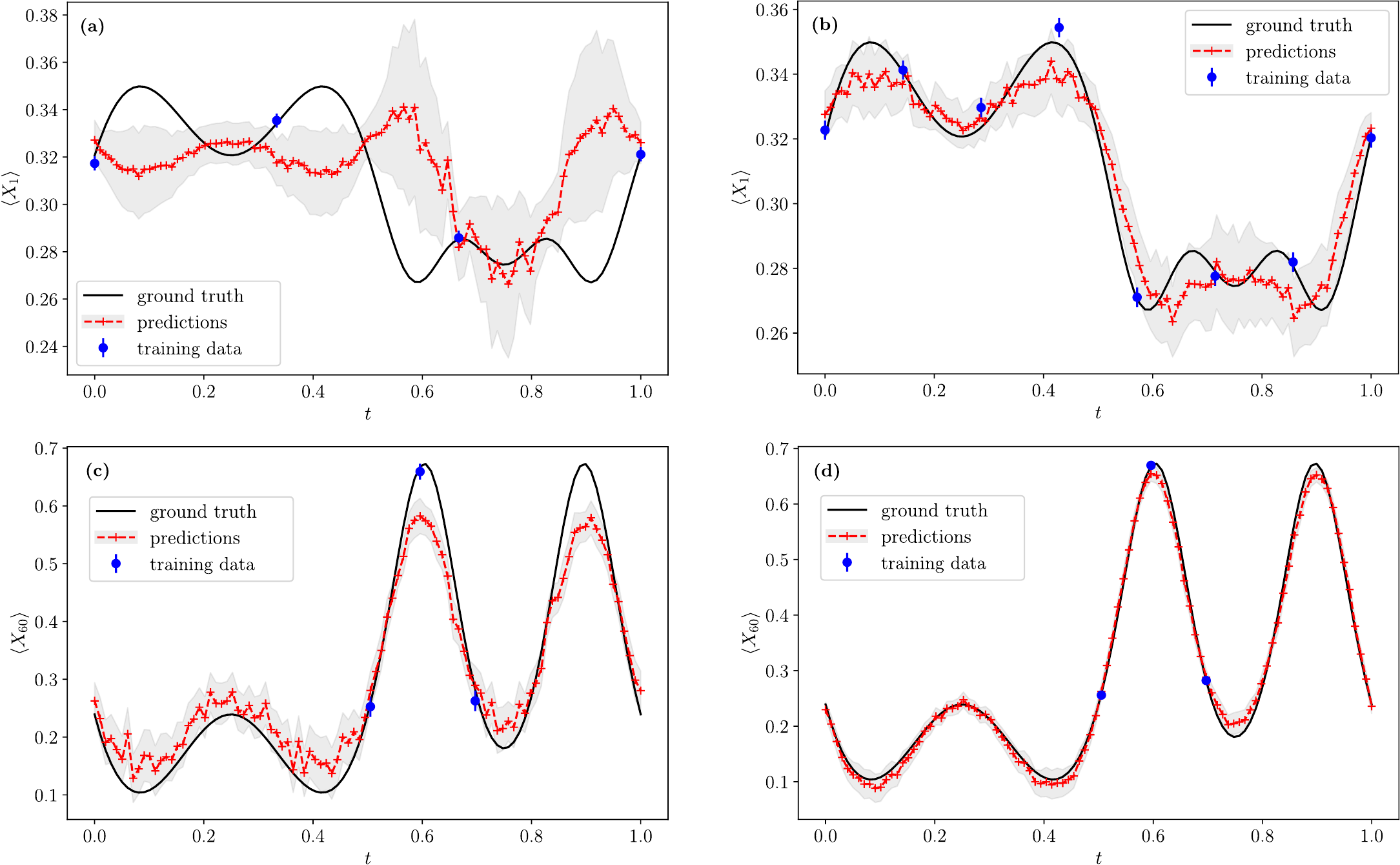}
\caption{{\bf QGP regression with the channel kernel for low and moderate shot budgets. }  We learn an expected value of  $X_1$ ($X_{60}$) versus an input state parameter $t$ for the setups  from Fig.~\ref{fig:GPR_Haar}.  In  (a, b), we plot QGP learning for the $n=4$ case, and shot budgets required for a prediction $N^{(p)}_s=1.4\cdot 10^6$ , $N^{(p)}_s=4.4\cdot 10^6$. The black curves are  ground truth, the blue circles are QGP training data, the red pluses are QGP predictions, and the shaded regions denote the 95\% prediction confidence interval.  In (c) and (d) we plot the results for the local channel ($n=64$) case, for  $N_s^{(p)}=2.7\cdot 10^4$ and  $N_s^{(p)}=9.0\cdot 10^5$, respectively. As in Fig.~\ref{fig:GPR_Haar}, we choose  QGP learning instances  with median RMSE in a set of 100 learning instances.  The training data error bars  represent their standard deviations of the mean. }
\label{fig:GPR_low_shots}
\end{figure*}

\subsection{Gradient descent baseline for Bayesian optimization demonstration}
\label{app:gd}

We perform 9 GD runs,  with the same initial points as for the BO runs, specified in Eq.~\eqref{eq:bo_gd_initial_grid_appendix}.  For each initial point $\vec{t}_0\in\mathcal{I}_0$ we generate a trajectory  of $N_{\rm GD}$ points defined as 
\begin{equation}
    \vec{t}_{r+1}
    =
    \Pi_{\Omega}
    \left[
        \vec{t}_r-\eta_r\,\widehat{\nabla}\mathcal{L}(\vec{t}_r)
    \right]\,,
    \label{eq:projected_gd_appendix}
\end{equation}
where $\Pi_{\Omega}$ denotes the projection back onto the rectangle $\Omega$ in case the update were to push the parameters out of it, $\eta_r$ is the learning rate, and $\widehat{\nabla}\mathcal{L}$ is a finite-difference estimate of the gradient~\cite{nocedal2006numerical}. We use central differences,
\begin{align}
    \partial_{\theta}\mathcal{L}(\theta,q)
    &\approx
    \frac{
        \mathcal{L}(\theta+\epsilon_{\theta},q)
        -
        \mathcal{L}(\theta-\epsilon_{\theta},q)
    }{2\epsilon_{\theta}}\,,
    \\
    \partial_q\mathcal{L}(\theta,q)
    &\approx
    \frac{
        \mathcal{L}(\theta,q+\epsilon_q)
        -
        \mathcal{L}(\theta,q-\epsilon_q)
    }{2\epsilon_q}\,,
\end{align}
and set $\epsilon_\theta=\epsilon_q=0.02$, $N_{\rm GD}=200$ and $\eta_r=0.1$.
At the boundary of $\Omega$, one-sided finite differences are used.
Since a single GD iteration away from the boundary requires four loss evaluations to estimate both gradient components, a GD trajectory with $N_{\rm GD}$ updates typically costs
\begin{equation}
    C_{\rm GD}^{(1)}
    =
    4N_{\rm GD}
\end{equation}
loss evaluations.

\section{Supplementary QGP regression results for low and moderate shot budgets}
\label{app:GPR_low_shots}

In Fig.~\ref{fig:GPR_low_shots}, we show representative channel  QGP learning results for the $n=4$ and $n=64$ systems investigated in Section~\ref{sec:channel_GPR_numerics},  with $N_s^{(p)}$ of order $10^4$--$10^6$ which is lower than $N_s^{(p)}$ of order $10^7$--$10^8$ used in Fig.~\ref{fig:GPR_Haar}. We see that for $n=4$ we learn the most prominent feature of $\langle X_1 \rangle$ versus $t$ with $N^{(p)}_s=1.4\cdot 10^6$ and $k=4$, and for $N_{s}^{(p)}=4.4\cdot 10^6$ and $k=8$ we detect all  minima and maxima of $\langle X_1 \rangle(t)$ apart from the shallowest dip.
In the  local channel  case ($n=64$),  $N_s^{(p)}=2.7\cdot 10^4$ ($k=3$) is large enough to learn all the observable features versus $t$, while with $N_s^{(p)}=9.0\cdot 10^5$ we see significant improvement of the accuracy of the predicted $\langle X_{60} \rangle(t)$ in comparison to the lower shot budget.

\end{document}